\documentclass[sigconf,nonacm]{acmart}
\pdfoutput=1
\usepackage{listings}
\usepackage{subcaption}
\usepackage{placeins}
\usepackage{notoccite}

\makeatletter
\let\orig@lstnumber=\thelstnumber

\newcommand\lstresetnumber{\global\let\thelstnumber=\orig@lstnumber}
\makeatother

\begin{document}
\title{Supercharging the APGAS Programming Model with Relocatable Distributed Collections}

\author{Patrick Finnerty}
\orcid{0000-0002-9037-967X}
\affiliation{
  \department{Graduate School of System Informatics}
  \institution{Kobe University}
  \city{Kobe}
  \country{Japan}
}

\author{Yoshiki Kawanishi}
\orcid{0000-0002-1233-7773}
\affiliation{
  \department{Graduate School of System Informatics}
  \institution{Kobe University}
  \city{Kobe}
  \country{Japan}
}

\author{Tomio Kamada}
\orcid{0000-0002-1646-1683}
\affiliation{
  \department{Department of Intelligence and Informatics}
  \institution{Konan University}
  \city{Kobe}
  \country{Japan}
}

\author{Chikara Ohta}
\orcid{0000-0002-4143-9429}
\affiliation{
  \department{Graduate School of System Informatics}
  \institution{Kobe University}
  \city{Kobe}
  \country{Japan}
}

\begin{abstract}
In this article we present our relocatable distributed collections library. 
Building on top of the AGPAS for Java library, we provide a number of useful intra-node parallel patterns as well as the features necessary to support the distributed nature of the computation through clearly identified methods.
In particular, the transfer of distributed collections' entries between processes is supported via an integrated relocation system.
This enables dynamic load-balancing capabilities, making it possible for programs to adapt to uneven or evolving cluster performance.
The system we developed makes it possible to dynamically control the distribution and the data-flow of distributed programs through high-level abstractions.
Programmers using our library can therefore write complex distributed programs combining computation and communication phases through a consistent API.

We evaluate the performance of our library against two programs taken from well-known Java benchmark suites, demonstrating superior programmability, and obtaining better performance on one benchmark and reasonable overhead on the second.
Finally, we demonstrate the ease and benefits of load-balancing and on a more complex application which uses the various features of our library extensively.
\end{abstract}


\maketitle

\section{Introduction}

Modern supercomputers that rely on many-core processors provide a large level of parallelism, both within a node and between nodes.
On the other hand, Beowulf clusters can provide a flexible environment for smaller-scale computations.
However the performance of distributed programs may suffer if potential disparities in the hardware used are not addressed.
Writing programs that can execute efficiently on either such systems and achieve good performance on both is therefore a challenge.

Writing parallel and distributed programs is inherently difficult, with many dedicated languages, runtime libraries, and programming models attempting to reduce the difficulty by introducing abstractions to programmers.
MPI~\cite{mpi} defines a standard for communication between processes but it does not provide support for intra-process parallelism, forcing programmers to rely on other libraries such as OpenMP in Hybrid MPI approaches~\cite{mpiplusx, hybridmpi, hybrydmpibrain}.
While such approaches allow for high-performance applications to be developed, they require programmers to become experts in multiple programming models and libraries.

The Partitionned Global Address Space (PGAS) programming model introduces elements that allow programmers to grasp the distributed nature of their program directly from the language used.
Several languages adopt this programming model, including UPC, Chapel, Habanero-Java~\cite{upc,chapel,chapel2,chapel3,habanero}.
UPC supports distribution of arrays across processes, giving the illusion of a shared-memory environment to programmers using ``shared'' pointers that live in the global address space.
It is however not possible to choose a custom distribution or to dynamically modify the distribution after array creation.
Chapel also allows programmers to select cyclic, block-cyclic distributions for their collection, but similarly does not allow subsequent changes to this distribution.

On the other hand, Charm++~\cite{charmpp,charmppcompared} proposes a unified programming model for parallel and distributed computation. 
The \emph{chare} abstraction is used to represent a relocatable object processing unit, with ``messages'' sent to and from chares representing a remote procedure call.
Load-balancing is done using by the Charm++ runtime following pre-implemented strategies, surrendering the distribution control to the system.
This makes developing location-aware programs difficult.
Also, the order into which messages are processed is non-deterministic, which can cause difficulties if the program needs to produce deterministic results.

Our objective is to allow programmers to manage the entry distribution of collections both explicitly and dynamically. 
We aim at providing support for common computation and communication patterns on these distributed collections as well common parallel patterns within a host.
To this effect, we introduce in this article our \emph{relocatable distributed collections} library.

Relying on a combination of the APGAS for Java programming model~\cite{apgasforjava} and MPI, our library makes it possible to write complex distributed and parallel programs with ease.
The distribution management of entries in our distributed collections is explicit, making it possible for programmers to freely re-organize entries over the course of the program execution through high-level abstractions.
We introduce the notion of ``teamed operation'' to describe computation or communication patterns that involve multiple processes.
We also propose a number of intra-node parallel patterns, such as reductions and producer/receiver.
Our distributed collections come with an API close to that of the Java standard library, providing a sense of familiarity to programmers who are then capable of reusing any prior knowledge.

To demonstrate the benefits of our library, we ported two programs from well-known benchmark suites~\cite{renaissance, javagrande}. 
We demonstrate superior programmability and performance one of them, and reasonable overhead on the second.
We also demonstrate the capability of a simulator featuring complex communication patterns to dynamically balanced its load on a Beowulf cluster featuring uneven performance across compute nodes thanks to the features of our library.

The remainder of this article is organized as follows.
We start by recalling some useful background in Section~\ref{sec:background}.
We then formally introduce key concepts of our relocatable distributed collections in Section~\ref{sec:collectionlibrary}.
In Section~\ref{sec:motivatingcases}, we showcase the main features of our library using actual distributed programs written with our library. 
We then discuss specific design choices and select implementation details in Section~\ref{sec:design}.
We evaluate the performance of our library in Section~\ref{sec:evaluation} before discussing related work in Section~\ref{sec:relatedwork}.
Finally, we conclude and discuss future work in Section~\ref{sec:conclusion}.

\section{Background}
\label{sec:background}

\subsection{APGAS for Java}
\label{sec:apgasjava}

The Partitioned Global Address Space (PGAS) is a programming model which brings specific constructs to handle locality to a programming language.
X10 further expands the PGAS programming model into Asynchronous PGAS (AGPAS) by providing support for asynchronous activities through dedicated keywords~\cite{x10}.
The APGAS for Java library~\cite{apgasforjava} mimics the keywords of X10 using static methods to bring the expressiveness of the model to the Java language.
With this library, Java effectively becomes an APGAS language.

An example ``Hello World'' program written with APGAS for Java is presented in Listing~\ref{lst:pureapgas}.
In APGAS for Java, class \verb+Place+ on line~2 is used to represent the locality and corresponds to a process running on a host.
The process allocation to physical computer is decided when launching a program, with a typical approach consisting of assigning one process (or ``Place'') per host.
Method \verb|asyncAt| is used to spawn an asynchronous activity on the place specified as parameter.
The variables and objects used in the asynchronous activity are automatically serialized to be transmitted to process they are run.
The \verb|finish| method on line~1 is used to wait until all asynchronous activities transitively spawned within its closure complete.
In the example shown in Listing~\ref{lst:pureapgas}, the main thread running on Place 0 will not progress further than the \verb|finish| method until all places have written their ``Hello'' message on the standard output.

\begin{lstlisting}[float,linewidth=\textwidth,caption=Distributed Hello World program with APGAS for Java,label={lst:pureapgas}]
finish(()->{
	for (Place p : places()) {
		asyncAt(p, ()-> {
			System.out.println("Hello from " + here()));
		}
	}
});
\end{lstlisting}

\subsection{Combining APGAS with MPI}
\label{sec:mpiinjava}

There are several projects that bring MPI to the Java programming language~\cite{mpijava, mpjexpress, openmpijava}, most often through a compatibility layer implemented between the Java program and the ``native'' C MPI calls using Java Native Interface.

The APGAS for Java and MPI runtimes are quite compatible. Each process becomes the combination of an APGAS ``Place'' and an MPI ``rank'' and the terms ``process,'' ``Place,'' and ``rank'' can therefore be used interchangeably in this context.
One difference when combining APGAS for Java and MPI is that unlike pure MPI programs, only rank/place 0 runs the program ``main''. Code is executed on other ranks through asynchronous activities managed by AGPAS.

\section{Relocatable Distributed Collections library}
\label{sec:collectionlibrary}

In this section we present the fundamental concepts introduced in our distributed collections library.
We start by introducing some supplements to the existing APGAS constructs in Section~\ref{sec:supplement}.
We then define what a distributed collection is in the context of the APGAS programming model in Section~\ref{sec:collection}.
We then present the notion of \emph{local handle} in Section~\ref{sec:localhandles}.
We then introduce the notion of \emph{teamed operation} and how our collections support intra-node parallelism in sections~\ref{sec:teamedoperations} and~\ref{sec:parallelism}.

Actual use-cases for our library will be presented in Section~\ref{sec:motivatingcases}, while design and implementation details are detailed in Section~\ref{sec:design}.

\subsection{Supplement to the APGAS for Java Library}
\label{sec:supplement}

As part of our library, we introduced some classes that supplement the existing APGAS constructs.
While these additions have little technical merit on their own, they bring some convenience to the programming model of APGAS and are use throughout our library.
The most significant addition for the purposes of our library consists in class \verb|TeamedPlaceGroup|.

Class \verb|TeamedPlaceGroup| represents a group of APGAS places and proposes a \verb|broadcastFlat| method taking a closure as parameter.
This method spawns the provided closure in an asynchronous activity on each place within the group and returns when the provided closure has completed on all places.
A ``world'' group which contains all the places participating in the computation is initialized by our library and can be obtained through the \verb|TeamedPlaceGroup.getWorld()| method.
Other groups containing a subset of the ``world'' can be created at will.

\begin{lstlisting}[float,caption=Equivalent program to Listing~\ref{lst:pureapgas} using class TeamedPlaceGroup,label={lst:withlib}]
TeamedPlaceGroup world = TeamedPlaceGroup.
	getWorld();
world.broadcastFlat(()-> {
	System.out.println("Hello from " + here());
});
\end{lstlisting}

We introduce Listing~\ref{lst:withlib} to illustrate the benefit of using class \verb|TeamedPlaceGroup|.
Notice that the \verb|broadcastFlat| method call on line~3 replaces the finish/asyncAt loop used in Listing~\ref{lst:pureapgas}.
Overall it is a practical shorthand which simplifies programs by mimicking the MPI programming style within a clearly identified block.
We use it extensively when writing programs with our library.
Internally, it carries an MPI communicator which is used by our library to communicate information between the places participating in the group.
A number of convenience methods that translate APGAS places into MPI ranks and vice-versa are also provided.

\subsection{Relocatable Distributed Collections}
\label{sec:collection}

\begin{table*}[h]
	\caption{Collection classes proposed by our library}
	\label{tbl:collections}
	\begin{tabular}{ll}
		\hline
		Collection & Description \\
		\hline
		\verb|Bag<T>| & Iterable set \\
		\verb|DistBag<T>| & Distributed variant of class \verb|Bag| \\
		\verb|CachableArray<T>| & Array used to share and replicate information across processes \\
		\verb|ChunkedList<T>| & Arbitrary long-index array \\
		\verb|DistChunkedList<T>| & Distributed variant of \verb|ChunkedList| \\
		\verb|DistCol<T>| & Variant of \verb|DistChunkedList| whose distribution is tracked \\
		\verb|CachableChunkedList<T> | & Variant of \verb|DistCol| whose entries can be replicated on multiple hosts \\
		\verb|DistMap<K,V>| & Distributed map from \verb|K| to \verb|V| objects \\
		\verb|DistConcurrentMap<K,V>| & Variant of \verb|DistMap| with additional protections against concurrency \\
		\verb|DistIdMap<V>| & Distributed map from \verb|long| indices to \verb|V| objects, its distribution is tracked \\
		\verb|DistMultiMap<K,V>| & Distributed map from \verb|K| objects to multiple \verb|V| objects \\
	\end{tabular}
\end{table*}

In the context of the APGAS programming model, a \emph{distributed collection} consists in a group of \emph{local handles} linked by a globally unique identifier.
We say that a collection is \emph{defined} on a group of places to represent the fact that a collection has a handle on each place belonging to this group.
When creating a new distributed collection, the \verb|TeamedPlaceGroup| on which the collection will be defined is given to the constructor as a parameter.
The main collections we provide with our library are summarized in Table~\ref{tbl:collections}.

\subsubsection*{Bag<T>}

The \verb|Bag| collection (and its distributed variant \verb|DistBag|) consist in a (distributed) iterable set.
Duplicated entries are allowed.
Special care was taken to its internal structure for it to efficiently receive elements from multiple concurrent threads.

\subsubsection*{CachableArray<T>}

A cachable array takes the form of an array containing objects that need to be replicated on each host and may be periodically updated.
Custom serialization and deserialization methods can be specified to use a user-chosen object to transport the updates to replicas.

\subsubsection*{ChunkedList<T>}

Class \verb|ChunkedList| and its distributed variants propose a collection which handles elements in multiple one dimension arrays mapped from ranges of \verb|long| indices.
We call each of these arrays mapped from a range of indices a ``chunk''.
Individual elements can be accessed and set through their \verb|long| index.
Some computations and/or manipulations on the distributed collections can be applied on ranges of entries.

We developed variants based on this class allow for more specific behaviors such as guaranteeing that chunks are unique across all hosts, or for chunks to be replicated on other hosts.
This enables support of various distributed applications in which replication of entries, entry distribution tracking, or other features are desired.

\subsubsection*{DistMap<K,V>}

The distributed map \verb|DistMap| is a generic distributed map taking \verb|K| objects as keys and \verb|V| objects as values.
\verb|DistMultiMap| is similar, but allows for multiple values to be mapped to a single key.

Before diving into specific features of our library let us first illustrate the notion of \emph{local handle} and \emph{teamed operation} with the sample program of Listing~\ref{lst:distmapexample} and the accompanying Figure~\ref{fig:distmapexample}.

\subsection{Local handle of a distributed collection}
\label{sec:localhandles}

\begin{lstlisting}[float,caption={Distributed map creation, record insertion, and relocation example},label={lst:distmapexample}]
TeamedPlaceGroup world = TeamedPlaceGroup.
	getWorld();
DistMap<String, String> dMap = 
	new DistMap<>(world);
dMap.put("main", "running");
world.broadcastFlat(()->{
  dMap.put(here(), "says hello");
  CollectiveMoveManager mm = 
  	new CollectiveMoveManager(world);
  if (here() == place(0)) {
    dMap.moveAtSync("main", place(1), mm);
  }
  mm.sync();
});
\end{lstlisting}

\begin{figure}
\begin{subfigure}[h]{\columnwidth}
\includegraphics[width=\textwidth,trim=0 3.5cm 16cm 6.7cm,clip]{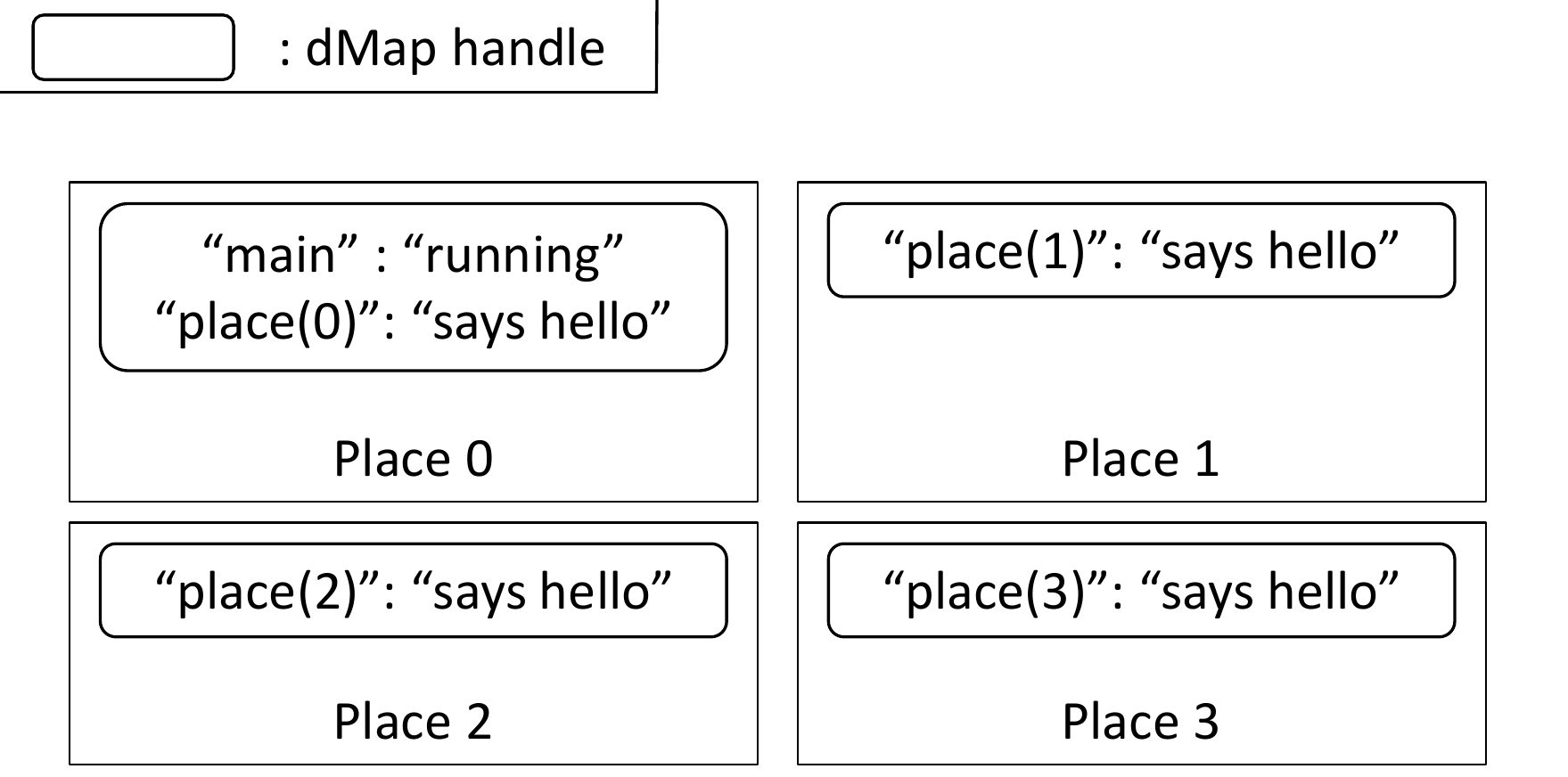}
\caption{before entry relocation}
\label{fig:distmapbefore}
\Description{This figure shows the state of entries on each of the 4 places during an execution of the program of Listing~\ref{lst:distmap} up until the entry relocation. Here follows the entries recorded in the local dMap handle of each Place:

  Place0:
    ``main'' --> ``running''
    ``place(0)'' --> ``says hello''

  Place1:
    ``place(1)'' --> ``says hello''

  Place2:
    ``place(2)'' --> ``says hello''

  Place3:
    ``place(3)'' --> ``says hello''

}
\end{subfigure}

\begin{subfigure}[h]{\columnwidth}
  \includegraphics[width=\textwidth,trim=0 3.5cm 16cm 5cm,clip]{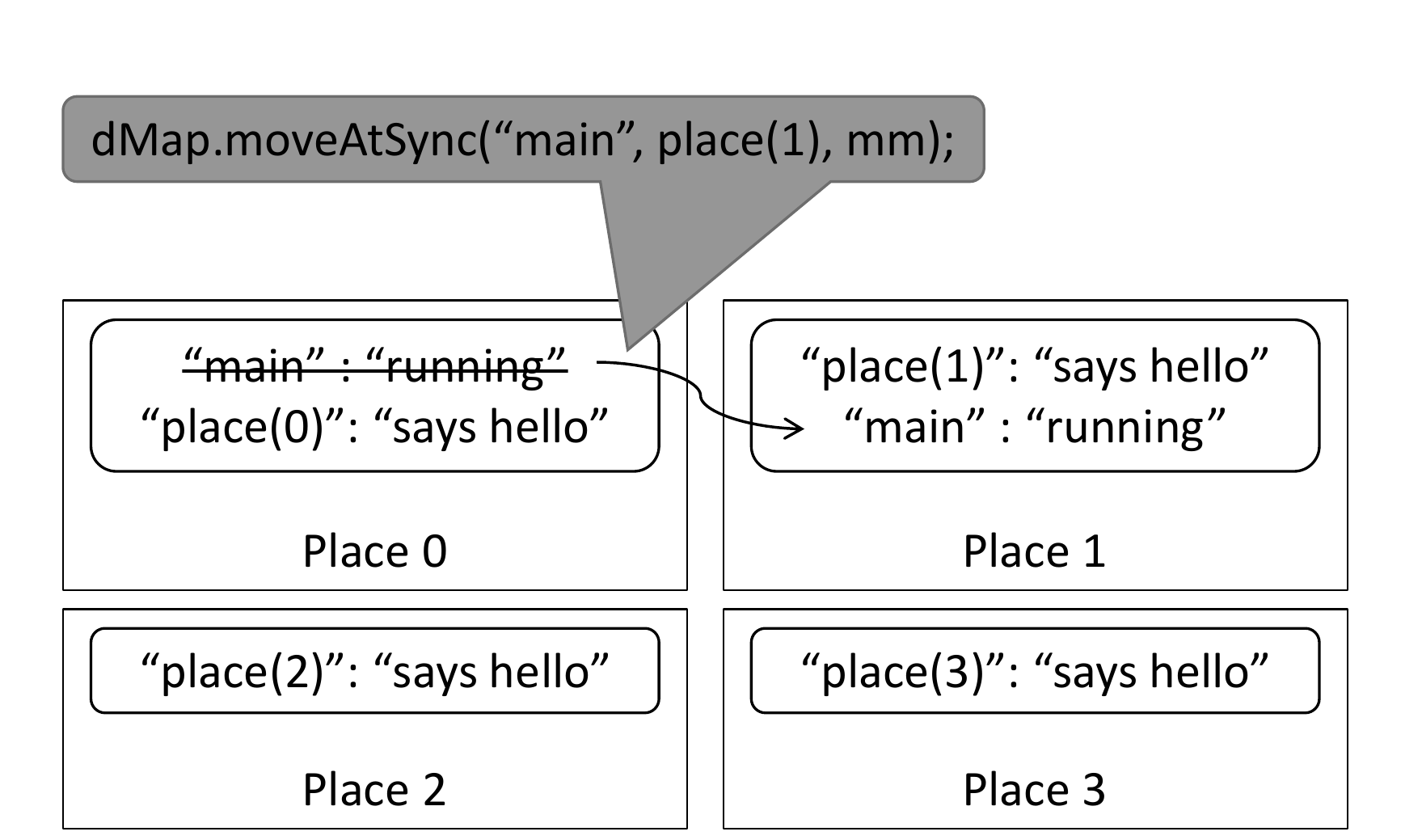}
  \caption{after entry relocation}
  \label{fig:distmapafter}
  \Description{This figure shows the state of entries on each of the 4 places after an execution of the program of Listing~\ref{lst:distmap}. Here follows the entries recorded in the local dMap handle of each Place:

  Place0:
    ``main'' --> ``running'' has been removed (relocated to Place1)
    ``place(0)'' --> ``says hello''

  Place1:
    ``place(1)'' --> ``says hello''
    ``main'' --> ``running'' has been received from Place0

  Place2:
    ``place(2)'' --> ``says hello''

  Place3:
    ``place(3)'' --> ``says hello''
  }
\end{subfigure}
\caption{State of the distributed map ``dMap'' in a  4 processes execution of the Listing~\ref{lst:distmapexample} program}
\label{fig:distmapexample}
\end{figure}

In Listing~\ref{lst:distmapexample}, a distributed map using \verb|String| for both keys and values is created on line~3.
This distributed collection is defined on the entire ``world,'' i.e. it will have a local handle on every process participating in the computation.
Then, a first entry is inserted on the running process on line~5.
The call to method \verb|put| only acts on the local handle registered on this place.
As such, the ``main'':``running'' entry is only be registered on the \emph{Place 0} handle.

On line~7, a second call to method \verb|put| registers new entries into the distributed map.
In this case however, since the call is contained in a \verb|broadcastFlat| method call, every place adds a new entry to their local handle.
Contrary to ordinary objects, the distributed collections used inside a closure are not copied to the remote processes but instead allocated on the fly.
As a result, the \verb|dmap| handles on Place 1-3 do not contain the entry previously placed in the handle of Place~0.
We provide more details about this topic in Section~\ref{sec:lazyalloc}.

Note that the key used to place new entries on line~7 differs on each host due to the APGAS method call \verb|here()| which returns the \verb|Place| object representing the currently running process.
Each local handle therefore contains a different key (``place(0),'' ``place(1)'' etc.) mapped to the \verb|String| \verb|''says hello''|, as is reflected in the contents of each local handle of \verb|dmap| in Figure~\ref{fig:distmapbefore}.

This illustrates the fact that conforming to the AGPAS programming model, all accesses to our distributed collections are ``local'' in the sense that APGAS asynchronous activities only ever interact with the handle of the distributed collection located on the process they are running.

\subsection{Teamed operations}
\label{sec:teamedoperations}

\emph{Teamed operation} is a generic term we use to describe operations or computations which involve some form of coordination or communication between the processes participating in the computation.
In Listing~\ref{lst:distmapexample} from line~8 onwards, we present one such teamed operation supported by our library in the form of an entry relocation between the handles of the \verb|distMap| distributed collection.

A ``collective move manager'' is first created on line~8.
This object is used to register entries of our distributed collections to be transferred from a handle to another.
In this case, only the first place decides to relocate the \verb|main:running| entry to Place~1, with all other places keeping their current entries.
The transfer is performed on line~13 when the \verb|mm.sync()| method is called by all the places participating in the computation.
The final state of the distributed map \verb|dmap| is what is presented in Figure~\ref{fig:distmapafter}.
In particular, note that the \verb|main:running| entry has been removed from the handle on Place~0 and inserted into the handle of Place~1.

There are a variety of ``teamed operations'' implemented in our library supporting various features, including reductions, entry relocation, replication etc.
We will introduce the most significant of them in the next section.
The key unifying characteristic of all our teamed operations is that they require the communication and synchronization between an asynchronous activity from each Place within a certain place group.

In the example presented above, the group of processes participating in a teamed relocation is determined by the \verb|TeamedPlaceGroup| object passed to the constructor of the \emph{collective move manager} on line~9.
Here, the \verb|world| place group is used, meaning that every place in the computation needs an asynchronous activity to call \verb|mm.sync()| before they can respectively resume their progress even if they do not send/receive any entry as part of the collective relocation.

Teamed operations pair nicely with the \verb|broadcastFlat| method of class \verb|TeamedPlaceGroup|, whose purpose is precisely to launch an asynchronous activity on each place of an identified group.
There is however no requirement to call ``teamed operations'' from within a matching \verb|broadcastFlat|.
This gives more experienced programmers the freedom to implement complex synchronization patterns by combining the usual finish/async constructs of APGAS for Java with the teamed operations proposed by our library.
For instance, if we wanted to allow Place~2 and Place~3 to continue their progress while Place~0 and Place~1 exchange entries, a different \verb|TeamedPlaceGroup| containing only the first two places could be used when creating the collective relocator on line~8, with only Place~0 and Place~1 calling the \verb|mm.sync()| method of that relocator on line~13.

\subsection{Support for intra-node parallelism}
\label{sec:parallelism}

As we will demonstrate in the next section, all our distributed collections feature typical \verb|forEach|, \verb|reduce| and other such methods that take a closure as argument. 
This closure is then applied to the entries contained in the local handle of the distributed collection.
Parallel variations of these methods are also implemented, allowing programmers to benefit from a multithreaded runtime without having to manually schedule the required threads.

Internally, we rely on the APGAS finish/async pair of constructs to spawn and control the threads needed for the parallel variants of these methods. 
For \verb|ChunkedList| and its variants, we allocate entries evenly between the threads available on the local host.
This is made trivial by the nature of this collection whose entries are recorded by ranges.

Spawning explicit activities on the library side also helps when objects dedicated to a single thread are needed by the computation pattern. 
This is the case for instance of the parallel producer/receiver pattern, reductions, and ``accumulators," presented in Section~\ref{sec:producerreceiver}, Section~\ref{sec:reduction}, and Section~\ref{sec:moldynaccumulator} respectively.

In each of these computation patterns, our library handles to allocation of the necessary objects for the threads to work in isolation from one-another.
This lightens the burden on programmers, re-focusing the program on the computation at hand rather than the schedule needed to support intra-node parallelism. 

\section{Motivating Cases}
\label{sec:motivatingcases}

In this section, we develop the abstractions available to programmers using examples taken from distributed programs written with our library.
We rely on the distributed implementation of the \emph{PlhamJ} financial market simulator, a distributed \emph{K-Means} implementation, and the N-body simulation \emph{MolDyn}.

Following a brief presentation of each program, we illustrate the abstractions and features they rely on in dedicated subsections.
The features are presented in order of appearance in their respective applications, but the reader may choose to forego this order and browse by feature category: \emph{intra-node parallelism}, \emph{teamed relocation}, and \emph{replication}.

\subsection*{PlhamJ}
\label{sec:plhamj}

\begin{figure*}
	\includegraphics[width=\textwidth,trim={0 0 0 0},clip]{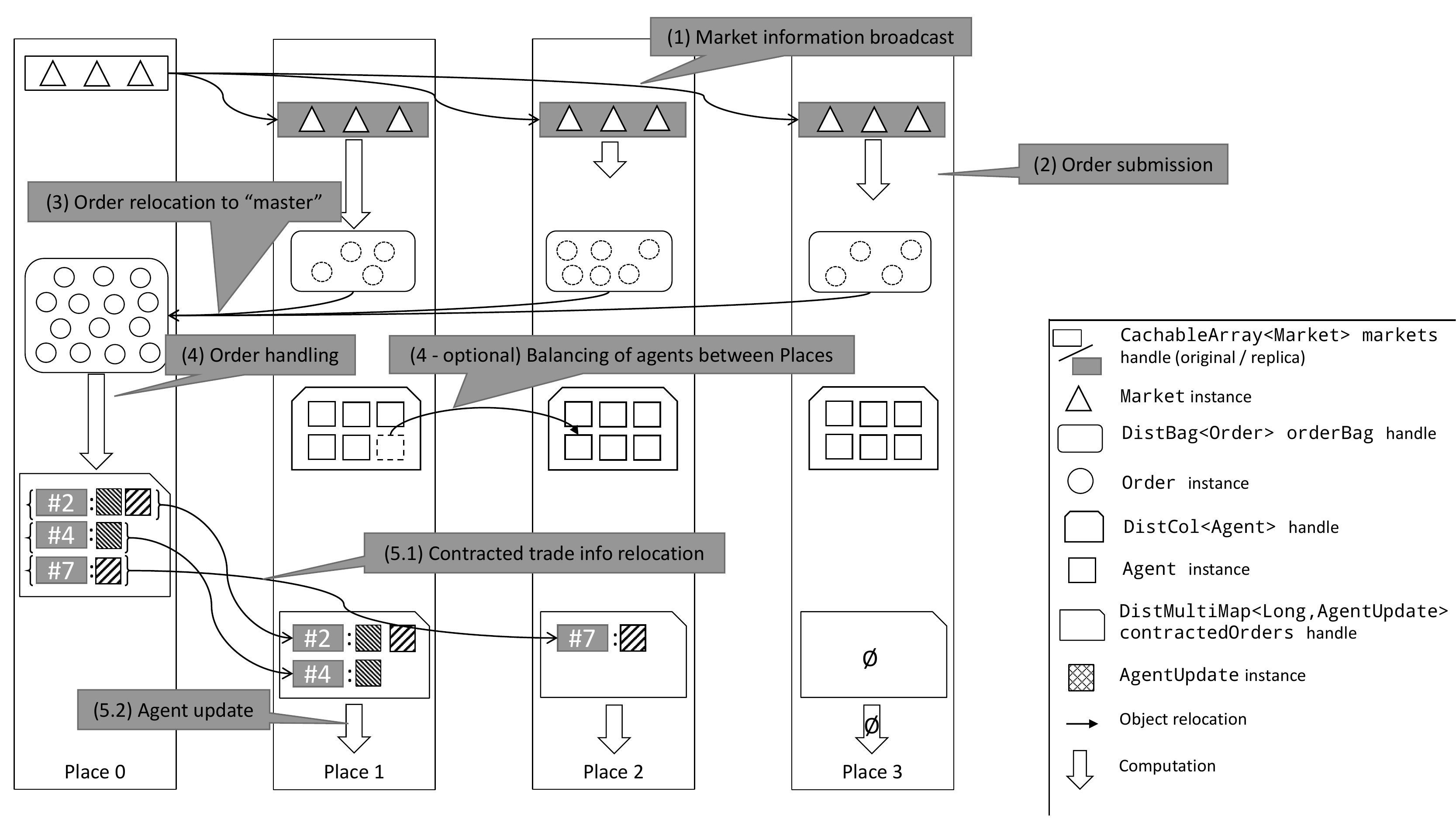}
	\caption{Figurative representation of the communications and computations processes that take place during a round of the Plham simulation}
	\label{fig:plhamiteration}
	\Description{This is a figurative representation of the various communication phases and computation that occurs during a round of the Plham simulator. Its contents are already discussed in part in the previous paragraph, but in this description, further details will be included.
		
		At the beginning of the round (step 1), the Market information is broadcast from the ``original'' handle located on Place~0 to its replicas on Place~1, Place~2, and Place~3.
		Then, the first computation phase (step 2) takes place. Agents held by Place~1 to Place~3 place the orders they want to make on the markets in the handle of distributed collection orderBag.
		
		In this example a total of~14 orders are submitted by the agents during step (2), with 4 orders in the orderBag handle of Place~1, 6 in the orderBag handle of Place~2, and 4 in the handleBag of Place~3.
		These 14 instances of class Order are then relocated into the orderBag handle of Place~0 as part of (step 3).
		
		The order-handling of step(4) results in~2 trades to be contracted: one trade between agent \#2 and agent \#4, and a second trade between agent \#2 and \#7.
		This results into two paris of updates to be made on the agents participating in the simulation.
		At this stage, there are therefore 3 entries into the contractedOrders multimap handle of Place~0: 2 AgentUpdate instannces mapped from key ``2'', 1 AgentUpdate mapped from key ``4'', and 1 AgentUpdate mapped from key ``7''.
		
		In this example, we assume that both agent~\#2 and~\#4 are located on place~1, while place~2 holds agent~\#7.
		The entries of the contractedOrders map are relocated according to this distribution in step (5.1) .
		After the relocation, the contractedOrders handle of Place~0 is empty.
		The handle of Place~1 contains the mapping from key ``2'' to the two updates for Agent~\#2 and the mapping from key ``4'' to the single update for Agent~\#4.
		The handle of Place~2 contains the mapping from key ``7'' to the single update of Agent~\#7.
		Place~3 holds no agents that were able to make a trade in this round. Therefore, there are no contracted trades contained in its local handle of \verb|contractedOrders| after the relocation.
		
		For the final step (5.2) consisting of updating the agents that contracted a trade, Place~1 updates agent~\#2 and agent~\#4, Place~2 updates agent~\#7.
		Place~0 does not make any update as it never contains any agent.
		Place~3 has no update to make as no agent it holds were able to make a trade in this round.}
\end{figure*}

\emph{Plham} is a financial market simulator first implemented in X10~\cite{plham}.
Simulations are prepared using a JSON configuration file which details the agents, markets, sessions, and events that will occur during the simulation.
Users of this program can prepare trader implementations by extending the included \verb|Agent| class.
The simulator produces configurable outputs based on the information available over the course of the simulation, with the produced results deterministic following an initial seed.
Internally several ``runners'' implementations are available (sequential, parallel ...).

A round of the Plham simulator comprises the following steps. First, agents place orders based on the current market information. Secondly, buy and sell orders placed by agents are matched to contract trades, updating the state of the market. Lastly, agents that have contracted a trade during this round are informed. These steps then repeat using the updated state of the markets for as many rounds as specified in the simulation configuration.

To make use of larger-scale computer clusters, a distributed version of the Plham simulator is available.
In this implementation traders are distributed over multiple processes to leverage the greater parallelism of the underlying distributed runtime.
However, this poses a number of challenges as the computation in charge of matching buy and sell orders needs to remain centralized on a single process (arbitrarily the first process, place~0) to provide the opportunity for high-frequency traders to place orders based on the most up-to-date market information.

As a consequence, we need to:
\begin{itemize}
\item propagate the updated state of the market to all the hosts participating in the simulation
\item relocate the \verb|Order| objects placed by agents to the centralized order-processing host
\item dispatch the contracted trade notifications to the processes that hold the intended Agent recipient
\end{itemize}

To further complicate matters, if one of the processes takes longer than the others to compute the orders of the agents it was assigned, the progress of the entire program is delayed.
In non-dedicated clusters, such a load unbalance can be caused by disparities in the hardware used to support the distributed computation (different CPUs, different frequencies or number of cores), or by other processes competing for resources.
This poses a challenge as it is not reasonable to create a specific initial distribution for each cluster and/or simulation.
Moreover, even ``ideal'' distributions would not be able to react to dynamic changes in the cluster' performance.
While we could implement dynamic load-balancing of agent across hosts to resolve these situations as they occur, this poses a problem when sending contracted trade information to agents as their location will evolve dynamically over time.

\emph{PlhamJ} is the Java implementation of Plham and was re-written using the features of our distributed collection library.
This gave us the opportunity to revisit the implementation of some communication patterns as well as integrating a simple dynamic load balancer within the simulator.
Under the distributed implementation of this simulator, a round takes place in 5 main computation and communication steps represented in Figure~\ref{fig:plhamiteration}:

\begin{enumerate}
  \item the updated state of the markets is broadcast to its replicas on the agent-handling processes
  \item the agent-handling processes collect the orders of the agents they hold
  \item these orders are gathered on the order-handling process
  \item the order-handling process tries to match sell orders with buy orders, creating an \verb|AgentUpdate| object for each agent involved in a trade. Meanwhile, the agents are balanced between the other processes so that they all take roughly the same time during the order submission step. In our load-balanced version, this is done every few rounds.
  \item the agent updates are dispatched to their respective Agent location (step 5.1) where the targeted agents are informed are then informed of the trades they made (step 5.2)
\end{enumerate}

In the following subsections, we detail with the accompanying code the various features of our library that support this implementation.

In an effort not to overwhelm the reader, we chose to introduce the relevant code piece-by-piece in each subsection.
Listing~\ref{lst:plhamjwholecode} in the Appendix consolidates all of them into a single Listing.

\subsection{Replication: CachableArray}
\label{sec:cachablearray}

In the Plham simulator, the most up-to-date market information is located in \verb|Market| objects located on the order-processing place. To replicate the updated state of the market information to the other processes in the computation, we rely on class \verb|CachableArray| as shown in Listing~\ref{lst:cachablearray}. 

\begin{lstlisting}[float,caption={Replication of Market objects in the PlhamJ simulator},label={lst:cachablearray}]
CachableArray<Market> markets;
world.broadcastFlat(() -> {
	// (1) Broadcast the updated state of markets
	markets.broadcast(MarketUpdate::pack,
		MarketUpdate::unpack);		
});
\end{lstlisting}

The replicas on the other processes are updated using the teamed operation \verb|broadcast| of line~15.
This method is called by all hosts participating in the computation and also serves as a synchronizing mechanism between the asynchronous activities running the simulation on each host.

The two methods given as parameter to this function, \verb|pack| and \verb|unpack|, are respectively used to extract information from the market objects and record it into a \verb|MarketUpdate| object, and to update the market replica based on the information contained in the \verb|MarketUpdate| object .
This allows the user to choose any object to carry the data necessary to update the objects. 

\subsection{Intra-node parallelism: Producer / Receiver}
\label{sec:producerreceiver}

\begin{lstlisting}[float,caption={Parallel Order collection and relocation in the Plham simulator},label={lst:orderproduction}]
DistCol<Agent> agents;
DistBag<List<Order>> orderBag;
world.broadcastFlat(() -> {
	// (2) Submit agent orders
	if (!isMaster) agents.parallelToBag(
		(agent, orderCollector) -> { 
			List<Order> orders = agent.
				submitOrders(markets);
			if (orders != null && !orders.isEmpty()) {
				orderCollector.accept(orders);
			}
		}, orderBag);
	
	// (3) Collect all orders on the ''master''
	orderBag.team().gather(place(0));
});
\end{lstlisting}

In the second step of a PlhamJ iteration, every agent is asked to submit its orders based on the current Market information.
This consists in calling method \verb|submitOrders| on every \verb|Agent| object participating in the computation.
This method returns a list of orders, with agents able to place a single, multiple, or no orders at all.
In Figure~\ref{fig:plhamiteration}, we represented a total of~14 orders submitted by the agents during step (2).
The order are collected into the \verb|DistBag| ``orderBag''.

The corresponding code is shown in Listing~\ref{lst:orderproduction}.
The method \verb|parallelToBag| called on line~5 relies on the internal features of class \verb|DistBag| to allow multiple threads to concurrently place the orders into the local handle of this collection. 
This method takes two parameters. 
The first one is a closure taking an Agent and an ``orderCollector'' as parameter.
This closure will be applied to every agent in the local \verb|agents| handle in parallel, with the ``orderCollector'' taking the value \verb|Bag| instance being used to collect the orders.
The second parameter to method \verb|parallelToBag| is the \verb|Bag| into which all collected objects will be placed.

In this particular case, empty or \verb|null| lists returned by agents that choose not to place any new order for this round are discarded using the condition on line~9.
In cases where every entry in the collection produces an object to record in the specified bag, more simple signatures of the \verb|parallelToBag| method can be used.

\subsection{Teamed relocation: Gather}

After each place has gathered the orders placed by its agents, all the orders are relocated to the order-processing place where they will matched to create trades.

This is performed when each host calls the \verb|gather| method of class \verb|DistBag|, as shown on line~15 of Listing~\ref{lst:orderproduction}.
This method is a teamed operation which needs to be called on every handle of the distributed collection \verb|orderBag| for the calling activities to progress. 
As such, it is used as a synchronization point between place~0 (which does not produce orders during the second step) and the other agent-processing places.

When the relocation has completed and all the orders produced during this rounds have been relocated to place~0, the order-matching computation of step (4) begins on Place~0. 

\subsection{Teamed relocation: Dispatch}
\label{sec:dispatch}

During the order-handling process, each trade contracted results in two \verb|AgentUpdate| objects to be created, one for each Agent involved in the trade.

In Figure~\ref{fig:plhamiteration} we show 2 trades to be contracted: one trade between agent \#2 and \#4, and a second trade between agent \#2 and \#7.
These agent updates are placed into the \verb|contractedOrders| distributed multi-map at the index matching their intended agent recipient.
In other words, if agent \#2 (contained in collection \verb|agents|) contracts a trade, the ``agent update'' containing this information is placed at index \#2 in the \verb|contractedOrders| handle of Place~0.

To inform the agents of the trades they contracted during this round, the entries of \verb|contractedOrders| first need to be relocated to the location of their intended recipient. 
This is done as part of step (5.1) where the current distribution of collection \verb|agents| is used to determine the new location of each entry in the multi-map.

The corresponding code is shown in Listing~\ref{lst:dispatch}.
First, the current distribution of \verb|agents| is retrieved on line~6.
This is possible thanks to the distribution tracking mechanism integrated in class \verb|DistCol| which contains the agents participating in the simulation.
Then, the entries of collection \verb|contractedOrders| are relocated at the place where the corresponding agent is located by calling method \verb|relocate| on line~8.

\begin{lstlisting}[float,caption={Dispatch of contracted order updates and agent update},label={lst:dispatch}]
DistCol<Agent> agents;
DistMultiMap<Long, AgentUpdate> contractedOrders;
world.broadcastFlat(() -> {
	// (5) Inform the agents of the trades they made
	// (5.1) Relocate contracted trade information
	LongRangeDistribution agentDistribution = 
		agents.getDistribution();
	contractedOrders.relocate(agentDistribution);
	// (5.2) Update the agents
	if (!isMaster) contractedOrders.parallelForEach(
		(idx, updates) -> {
			// Retrieve the agent targeted by the update
			Agent a = agents.get(idx);
			// Apply each update for this agent
			for (AgentUpdate u : updates) {
				a.executeUpdate(u);
			}
	});
});
\end{lstlisting}

This method is a teamed operation which relocates the entries it contains to match the distribution given as parameter. 
In this particular example, the location of \verb|Agent| is recorded in a mapping from \emph{ranges of indices} to \verb|Place| objects. 
This distribution is assimilated as a distribution from \verb|long| indices to \verb|Place| object by class \verb|DistMultiMap| to determine the new location of each individual key recorded in \verb|contractedOrders|.

For the illustration purposes of Figure~\ref{fig:plhamiteration}, we assume that both agent~\#2 and~\#4 are located on place~1, while place~2 holds agent~\#7.
The entries of the contracted trade information are therefore relocated according to this distribution; \verb|contractedOrders| entries with key~\#2 and~\#4 are relocated to Place~1, and the entry with key~\#7 is relocated to Place~2.
Place~3 holds no agents that were able to make a trade in this round.

In step (5.2), each agent which contracted trades during the previous round receives its updates in parallel using a typical \verb|parallelForEach| shown from line~9 to~17 in Listing~\ref{lst:dispatch}.
The signature used here takes both the index (\verb|idx|) and the list of updates (\verb|updates|) contained in collection \verb|contractedOrders| as parameter.
This allows retrieval of the targeted agent instance on line~12 by calling \verb|agents.get(idx)|.

\subsection{Teamed relocation: Load-balancing}
\label{sec:periodiclb}

In the situation presented in Section~\ref{sec:producerreceiver}, the order submission of agents takes place in parallel on several processes.
Initially, agents are distributed evenly across processes.
However this may not be ideal as disparities in the hosts used or the presence of competing processes on said host may introduce imbalance in the cluster.

As a result, some hosts take longer than other to process the agents they hold, delaying progress of the entire simulation.
We represented this in Figure~\ref{fig:plhamiteration} by different arrow lengths in step~(2), with Place~1 taking longer than all the other hosts to complete the order submission.

\begin{lstlisting}[float,caption={Load Balance step in PlhamJ simulator},label={lst:lbplhamJ}]
DistCol<Agent> agents;
long accumulatedOrderComputeTime = 0l;
int lbPeriod = 10; // load-balance period
int iter;          // current iteration number

world.broadcastFlat(() -> {
	finish(()->{
		// (4 - opt) balance agents between places
		if (iter % lbPeriod == 0) async(()->{
			// Exchange time information between hosts
			CollectiveMoveManager mm = 
				new CollectiveMoveManager(world);
			long [] computationTime = world.
				allGather1(accumulatedOrderComputeTime);	
			performLoadBalance(computationTimes, mm);
			mm.sync();
			accumulatedOrderComputeTime = 0l;
			agents.updateDist();
		});
		
		if (isMaster) handleOrders();
	});
});
\end{lstlisting}

Fortunately, our relocatable distributed collection library allows us to take measures when such a case occurs.
In the PlhamJ simulator, we introduced a load balancer mechanism shown in Listing~\ref{lst:lbplhamJ}.
On each host, the amount of time dedicated to computing the orders is accumulated into the local \verb|accumulatedOrderComputeTime| variable (not shown in previous listings, refer to line~17 and~22 of Listing~\ref{lst:plhamjwholecode} in the appendix).
After a chosen number of iterations have elapsed, the optional load-balancing step is triggered on line~9.
The load-balancing is performed in a dedicated asynchronous activity spawned using APGAS' \verb|async| method.
This means the load balancing (lines~9 to~19 in Listing~\ref{lst:lbplhamJ}) is done concurrently to the order-handling on Place0 (method \verb|handleOrders| on line~21).
The program progresses to the following step only when both the order handling and the load-balancing have completed thanks to the \verb|finish| of line~7 which contains both of these operations.

The transfer of agents is made using a \emph{collective relocator}, as was previously introduced in Section~\ref{sec:teamedoperations}.
To determine the number of agents to transfer, the processes first exchange the amount of time they each spent on the order-submission part of the main loop using a \verb|allGather1| call on lines~13-14.
This information serves as the basis for each host to decide if it gives agents away inside the \verb|performLoadBalance| method called on line~15.
In this method, the agent instances to relocate are registered into the collective relocator previously created on lines~11-12.

As a first approach, we chose to relocate agents from the most overloaded process to the most underloaded process.
We call this simplistic load-balancing strategy ``level-extremes''.
We will be able to revisit this part in later work to implement more sophisticated strategies.

The agents are then transferred between the handle of collection \verb|agents| when the teamed method \verb|sync| is called on line~16.
In Figure~\ref{fig:plhamiteration}, we represented this by one agent held by Place~1 being relocated to Place~2 to reflect the load-balancing decision based on previous iterations.
In reality, entire ranges of agents will be relocated, depending on how severely unbalanced the situation is.
The counter which tracks the time spent computing the agents' orders is then reset on line~17 so that the next load-balancing round takes information relevant to this new distribution.

We offer more details about the ways programmers can use to relocate entries of our distributed collections in Section~\ref{sec:relocation}.

\subsection{Distribution Tracking}

In the absence of an integrated entry location record, managing a distribution record manually comes with tremendous effort.
In essence, tracking the location of entries of a distributed collection requires the active maintenance of a second distributed collection, with each insertion, removal, and transfer of an entry in the first collection requiring an update into the second.
This would greatly obfuscate the code and increase the chances of introducing bugs into the program.

In our library, we have implemented the facilities that allow for tracking of entry location and relocation in two of our distributed collections, the distributed arbitrary index array \verb|DistCol|, and the distributed map \verb|DistIdMap|.
The premise of tracking the location of a distributed collection's entries implies that there exists some way to uniquely identify each entry.
In both of these collections, individual entries can be identified by their unique \verb|long| index.

Our distribution tracking system associates each index with the location (\verb|Place|) of the associated record.
However, in a concern for efficiency, we do not keep a location record for each individual index in the case of class \verb|DistCol|.
Instead, we rely on range descriptions of locations to reduce the number of key/value pairs necessary to record of the location of each entry in these distributed collection.

The information concerning entries relocated between handles, or entries added/removed from a handle is not eagerly propagated to the other handles of the distributed collection.
Instead our distribution management proposes a teamed update method through which the local distribution records of a collection are reconciled to reflect the actual distribution at the moment of the call.
We took care in the implementation of this process to only communicate the distribution changes that occurred since the previous \verb|updateDist| call in order to minimize the amount of information exchanged.
This is a teamed operation which consists in reconciling the distribution information contained in each handle of the distributed collection. 

In the PlhamJ simulator, we use class \verb|DistCol| to contain the agents participating in the computation.
It is the distribution tracking facilities of this class that allow us to dynamically relocate agents over the course of the simulation without compromising the dispatch of contracted trades update as was laid out in Section~\ref{sec:dispatch}.
After agents have been relocated, method \verb|updateDist| is called on line~18 of Listing~\ref{lst:lbplhamJ} to refresh the distribution information contained in each handle.
As a result, the distribution of agents obtained on line~6 of Listing~\ref{lst:dispatch} during the subsequent contracted trade dispatch will be up-to-date, guaranteeing that each agent involved in a trade receive their intended updates in step~(5) of the PlhamJ round.

\subsection*{K-Means}
\label{sec:kmeans}

K-Means is an iterative clustering algorithm which separates points into a pre-defined ``k'' number of clusters.
There are three steps in a K-Means iteration.
Starting with randomly selected initial centroids, each point is assigned to the cluster of its closest centroid.
Then, the average position of each cluster is computed.
Finally, the point closest to each average position is chosen as the new centroid for the next iteration.

We chose to adapt the K-Means algorithm from the Java Renaissance benchmark suite~\cite{renaissance}.
We rely on class \verb|DistChunkedList| to contain the points subject to the algorithm.
In this distributed version, each place participating in the computation takes care of the points it contains in its local handle.
Listing~\ref{lst:kmeans} presents the main computation loop of our distributed K-Means implementation.
The assignment of each point to a cluster is done in parallel using a \verb|parallelForEach| method call on line~12.
On the other hand, the average cluster position and the selection of the next centroid are implemented as \emph{teamed reductions} on lines~17 and 22 respectively.
We will discuss the implementation of a reducer and its embedded support for parallelism in Section~\ref{sec:reduction} first. Then, we will discuss the difference between a ``local'' reduction and the ``teamed'' reduction used in K-Means in Section~\ref{sec:teamedreduction}.

\begin{lstlisting}[float,caption={Distributed K-Means implementation with our collection library},label={lst:kmeans}]
TeamedPlaceGroup world = 
	TeamedPlaceGroup.getWorld();
DistChunkedList<Point> points;  // init' omitted
double[][] initialClusterCenter;// randomly chosen

world.broadcastFlat(() -> {
	double[][] clusterCentroids = 
		initialClusterCenter;
	for (int iter = 0; iter < repetitions; iter++) {
		final double[][] centroids = clusterCentroids;
		// Assign each point to a cluster
		points.parallelForEach(
			p -> p.assignCluster(centroids));

		// Compute the avg position of each cluster
		AveragePosition avgClusterPosition = 
			points.team().parallelReduce(
				new AveragePosition(K, DIMENSION));

		// Compute the new centroid of each cluster
		ClosestPoint newCentroids = 
			points.team().parallelReduce(
				new ClosestPoint(K, DIMENSION, 
					avgClusterPosition));

		// Update the centroids for the next iteration
		clusterCentroids = 
			newCentroids.closestPointCoordinates;
  }
});
\end{lstlisting}

\subsection{Intra-node parallelism: Reduction}
\label{sec:reduction}

To compute a reduction on the objects of one of our collection, a ``reducer'' object needs to be prepared.
This is the nature of classes \verb|AveragePosition| and \verb|ClosestPoint| which are used on line~16 and~21 of Listing~\ref{lst:kmeans}. These classes are in charge of computing the average cluster positions and the new centroids respectively.
Both of these classes are user-defined and extend the generic abstract class \verb|Reducer| provided by our library.

As part of a \verb|Reducer| implementation, programmers need to provide 3 methods:
\begin{itemize}
\item the \verb|newReducer| method which creates a new instance of the reducer
\item the \verb|reduce(T)| method which reduces the given \verb|T| object into this reducer instance
\item the \verb|merge(R)| method which merges the contents of the reducer given as parameter into this instance
\end{itemize}

When creating a custom reduction object, the programmer need not care about concurrency. 
Our library ensures that no reducer object is used concurrently by multiple threads.

When computing a parallel reduction, each thread participating in the computation is given its own dedicated reducer instance obtained through the \verb|newReducer()| method of the reducer object supplied as parameter.
Each thread then calls method \verb|reduce(T)| on the entries of the collection it was allocated with its dedicated reducer instance.
When all threads have reduced their attributed entries, the reducer objects are merged back into a single instance using method \verb|merge(R)| to obtain the final result.

\subsection{Teamed Reduction}
\label{sec:teamedreduction}

A \emph{local reduction} consists in a reduction computed on the entries contained in a single local handle.
A \emph{teamed reduction} on the other hand, is a reduction which is computed on all the entries contained in all the local handles of a distributed collection. 
They are accessible through a special \verb|team()| method to distinguish them from the reduction which operates on the local handle only.
In other words, method \verb|parallelReduce(R)| operates on the contents of the local handle of a distributed collection, while method \verb|team().parallelReduce(R)| used in the K-Means implementation shown in Listing~\ref{lst:kmeans} computes the reduction on the contents of the entire distributed collection.

A teamed reduction takes place in two stages.
First, a ``local'' reduction is computed following the process detailed in the preceding section.
Then, the local results of each handle are merged together into a single instance which is then returned as the result by each of the calling activities.
Internally, an MPI \verb|allReduce| call is made to communicate and compute the global result of the reduction across all running processes.
The MPI communicator used to make this call is the one of the \verb|TeamedPlaceGroup| on which the collection is defined.
The registration of the user-supplied reducer object necessary to use MPI object reductions is made automatically by our library.

The underlying use of MPI routines remains hidden from the user. 
The only practical consequence is that the teamed reduction call is blocking until all handles of the distributed collection complete their local reduction and exchange their results, after which each thread resumes its progress.

\subsection*{MolDyn}
\label{sec:moldyn}

MolDyn is a molecular simulation part of the Java Grande benchmark suite~\cite{javagrande} implemented with the MPI/Java compatibility layer MPJ~\cite{mpjexpress}.
It consists in a N-body simulation with all the force interaction between all the particles computed.
The particles are replicated on every host, with each host responsible for computing a subset of the force interactions.
This information is then communicated between all hosts before updating the position and velocity of each particle.

An iteration of the distributed MolDyn program takes place in three stages. 
First, a subset of the force interactions between the particles is computed on each host. 
Then, the force subjected to each particle are summed across hosts using an MPI allreduce call.
Finally, the position and velocity vectors of particles are updated.

We ported this benchmark using our distributed collections library to a hybrid implementation taking advantage of the multithreaded capabilities available within each process.
The arrays of \verb|double| used in the original implementation were converted to \verb|Particle| objects managed by a \verb|CachableChunkedList|.

Contrary to the previous examples we showed in this section, the computation pattern brought by MolDyn is no longer strictly ``owner-based''.
Instead of a particle operating based on its own information, it is the interaction between each pair of particles that serves as the basis for the computation. 
To support such patterns, we introduced class \verb|RangedListProduct|.
This class is used to represent combination pairs between the entries of two \verb|ChunkedList|s as depicted in Figure~\ref{fig:product} and provides a number of iterators and \verb|forEach| methods that act on the pairs it contains. 

As we did for PlhamJ, we will introduce the features needed to support this program piece by piece in the following subsections. 
The consolidated MolDyn program can be found in the appendix in Listing~\ref{lst:moldyn}.
The reader familiar with this benchmark will notice that the temperature scaling and the performance tracking are absent from the code we present here. 
These parts are included in our actual program, but we chose to omit them here to focus on the core part of the program. 

\subsection{Replication: CachableChunkedList}
\label{sec:moldynreplication}

We use the \verb|CachableChunkedList| distributed collection to contain the particles of the simulation.
Similar to the \verb|CachableArray| previously discussed in Section~\ref{sec:cachablearray}, this collections allows for entries to be replicated on multiple hosts. 
However, unlike the \verb|CachableArray|, \verb|CachableChunkedList| allows for multiple handles to be the primary owners of certain ranges of entries where the former only allows a single source to update the replicas. 

In the case of the MolDyn simulator, the particles are initialized on the first process in the distributed system.
At the start of the computation, these entries are replicated on the other hosts by calling the \verb|share| method on lines~8-12 in Listing~\ref{lst:moldynreplication}.
This teamed method takes one or multiple ranges as parameter and replicates the matching ranges of entries on the other hosts. 
In this particular case, only the first process shares the range of initialized particles on line~9, while the other processes (that do not contain any entries) merely receive the ranges shared by the other processes by calling the \verb|share| method without arguments on line~11. 

\begin{lstlisting}[float,caption={Particule replication in MolDyn},label={lst:moldynreplication}]
TeamedPlaceGroup world = 
	TeamedPlaceGroup.getWorld();
LongRange particleRange = 
	new LongRange(0, nbParticles);
CachableChunkedList<Particle> particles; 

world.broadcastFlat(() -> {
	if (world.rank() == 0) {
		particles.share(particleRange);	
	} else {
		particles.share();
	}
});
\end{lstlisting}

\subsection{Ranged List Product}
\label{sec:moldynproduct}

\begin{figure}
	\includegraphics[width=\columnwidth,trim={355 0 0 0},clip]{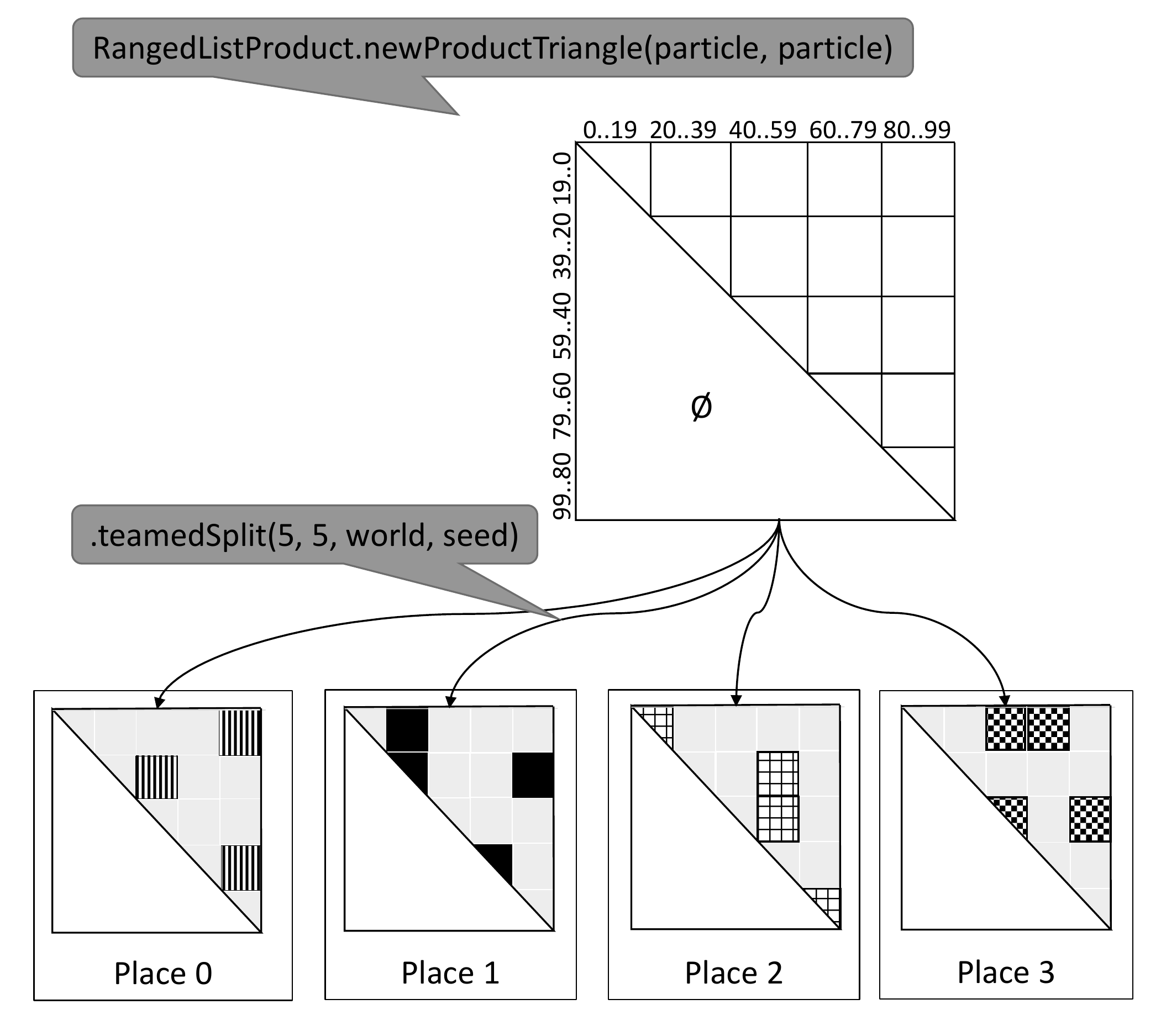}
	\caption{Illustration of the teamed split product used to represent the particle interaction pairs in our MolDyn implementation}
	\label{fig:product}
	\Description{}
\end{figure}

Creating a product between two ranged lists is done by calling a factory methods provided by class \verb|RangedListProduct|. 
In listing~\ref{lst:moldynrlp}, this is done on line~12 where the \verb|newProductTriangle| method is called.
The ranged list containing the particles is given as argument to this method as it takes the role of both operands. 
We note that this method eliminates the mirrored pairs as depicted in Figure~\ref{fig:product}: only the pairs residing in the upper triangle are included in the \verb|product| object.

In a second stage, the pairs of entries to process by each host are determined by calling the \verb|teamedSplit| method on line~14. 
This method performs two operations. 
First, it splits the pairs contained in the product into tiles, creating as many columns and lines as was specified as parameter.
If we assume that there are 100 particles in the simulation and that 5 columns and 5 rows are created, each tile will cover and area of 20x20 pairs, as depicted in Figure~\ref{fig:product}.

Then, a new instance of \verb|RangedListProduct| containing a subset of the created tiles is returned.
The \verb|TeamedPlaceGroup| given as parameter is used to determine the number of hosts involved in the ``split''. 
The running process' position inside the group and the seed are used to select the assignments returned by this method call.

Although not communication takes place, we still consider this operation to be ``teamed'' as it needs to be called with the same parameters on all processes participating in the computation to operate correctly.
This guarantees that every tile gets processed by at least one host as depicted in the lower part of Figure~\ref{fig:product}.

We note that the use of tiles in our implementation differs from the original MolDyn implementation where the rows of the upper triangle are allocated to each host in a cyclic manner.

\begin{lstlisting}[float,caption={Force interaction computation using RangedListProduct and Accumulators in MolDyn},label={lst:moldynrlp}]
TeamedPlaceGroup world = 
	TeamedPlaceGroup.getWorld();
LongRange particleRange = 
	new LongRange(0, nbParticles);
CachableChunkedList<Particle> particles;
int Ndivide = 5;
long seed = 0;

world.broadcastFlat(() -> {
	RangedList<Particle> prl = 
		particles.getChunk(particleRange);
	RangedListProduct<Particle, Particle> product = 
		RangedListProduct.newProductTriangle(prl,prl);
	product = product.teamedSplit(Ndivide, Ndivide,
		world, seed);
	
	Accumulator<Sp> acc = 
		new AccumulatorCompleteRange<>(
		particleRange, Sp::newSp);
	
	product.parallelForEachRow(acc,
		(Particle p, RangedList<Particle> pairs, tla)
			-> force(p, pairs, tla));
	
	particles.parallelAccept(acc,
		(Particle p, Sp a) -> p.addForce(a));
});
\end{lstlisting}

\subsection{Intra-node parallelism: Accumulator}
\label{sec:moldynaccumulator}

The conversion to an hybrid implementation which uses local parallelism to compute the force interaction between the particles brings about an additional challenge compared to the single-worker-per-host implementation of the Java Grande benchmark. 
In the original implementation, the force sum can be written directly to the particles. 
However in a hybrid implementation this is no longer possible as there would a risk that two threads concurrently write the contribution of interactions involving the same particle.
To address this issue, we introduced what we call \emph{accumulators} to our library.

This mechanism (no relation to the \verb|LongAccumulator| or the \verb|DoubleAccumulator| classes from the standard \verb|atomic| package) is used by threads participating in a parallel computation to store information independently from one-another.
The \verb|Accumulator| object serves as a factory for multiple ``thread-local accumulators'' which are objects dedicated to an individual thread during a parallel computation.
In turn, each of these ``thread-local accumulators'' will contain individual objects of any user-chosen type into which information can be stored at a specified index.
These individual objects are initialized using the function given as parameter at the time of the \verb|Accumulator| creation.

An accumulator's lifecycle takes place in 3 phases: (1) creation, (2) accumulation of information into the accumulator, and (3) acceptance of the accumulated information by an existing collection. 
In the case of MolDyn, the accumulator used during the force computation is created on line~17-19 of Listing~\ref{lst:moldynrlp}.
The type used to store information in regards to each particle is class \verb|Sp|, which contains 3 \verb|double| members to represent the ``x,y,z'' force components.

The force computation takes place on lines~21-23. 
Let us briefly detail what method \verb|parallelForEachRow| does. 
The closure it takes as parameter will be applied to each row of the tiles contained in the underlying \verb|RangedListProduct|. 
The first parameter of the closure \verb|Particle p| consist in the first half of the particle pairs to compute within this method, while the second half of the pairs are provided by the second \verb|RangedList<Particle> pairs| argument. 
Inside method \verb|force|, the force resulting of each interaction is stored into the \verb|Sp| instance dedicated to the involved particles.
The thread-dedicated \verb|Sp| instances are available through the third parameter of the closure: \verb|tla|.
This parameter is populated by our library using the \verb|acc| accumulator given as the first parameter to the \verb|parallelForEachRow| method on line~21. 

Finally, the information stored in the various \verb|Sp| objects is used to apply changes to the particles using the \verb|parallelAccept| method as demonstrated on line~25-26 of Listing~\ref{lst:moldynrlp}. 
The closure given as parameter to the \verb|parallelAccept| method sums the force vectors contained in the various \verb|Sp| instances into the dedicated member of the particles.
Internally, this closure is applied to each \verb|Sp| instances prepared for each thread that participated during in the ``accumulation'' phase. 

Here, we demonstrated the use of the accumulator for a single computation before using it to modify a collection.
It is also possible to perform multiple accumulations on various collections before ``accepting'' the accumulator. 

\subsection{Replication: Reduction}
\label{sec:moldynreduction}

After the force contribution computed on each host has been completed and integrated into the local replicas of each particule, the replicas all bear different force components due to the different subset of interactions that was computed on each host. 
To reconcile the force subjected to each particle, a reduction is made on each particle shared by the local handles of the \verb|CachableChunkedList| used to support the program.

This is done on lines~6 to~14 in Listing~\ref{lst:reduction} using method \verb|allreduce|. 
This is a specific feature of class \verb|CachableChunkedList| operating on the entries shared across hosts.
Unlike the teamed reduction discussed in the context of K-Means in Section~\ref{sec:teamedreduction}, in this situation each particle replica of matching indices are reduced into a single instance and stored back into the local handle of the \verb|particles| collection.

Contrary to previous examples of object relocation and replication, we demonstrate here the capabilities of our library to support primitive-type communication patterns.
In this case, the force information is converted from each particle into three \verb|double| using the first closure running from line~6 to line~9. 
Then the MPI operation \verb|MPI.SUM| is used to reduce these raw types. 
Finally, the reduced values are written back into the particle entries in each host using the second closure running from line~10 to~13.

Internally, buffer arrays of the appropriate length are automatically allocated based on the number of entries shared between hosts and the number of raw types used to describe each entry.
This allows for more efficient use of MPI functionalities as serializing the entire particle object and implementing a custom reduction on this object is not necessary here.

After the force subjected to each particle has been consolidated across all hosts, each particle ``moves'' (i.e. updates its position and velocity vector) on line~16 of Listing~\ref{lst:reduction}, concluding an iteration of the program. 

\begin{lstlisting}[float,caption={Force reduction on each particule in the MolDyn simulation},label={lst:reduction}]
TeamedPlaceGroup world = 
	TeamedPlaceGroup.getWorld();
CachableChunkedList<Particle> particles;

world.broadcastFlat(() -> {
	particles.allreduce((out, Particle p) -> {
		out.writeDouble(p.xforce);
		out.writeDouble(p.yforce);
		out.writeDouble(p.zforce);
	}, (in, Particle p) -> {
		p.xforce = in.readDouble();
		p.yforce = in.readDouble();
		p.zforce = in.readDouble();
	}, MPI.SUM);
	
	particles.parallelForEach(p -> move());
});
\end{lstlisting}

\section{Design \& Implementation}
\label{sec:design}

In this section, we detail select design elements and implementation topics of our distributed collection library that were not detailed in the preceding section.
We also briefly demonstrate how to compile and execute programs with our library.

\subsection{Lazy Allocation of Local Handles}
\label{sec:lazyalloc}

For every distributed collection whose classes we presented in Table~\ref{tbl:collections}, there is in reality one instance of the corresponding class on each process on which the collection is defined.
These instances implement what we refer to as the ``local'' handles of distributed collections.

When a distributed collection is created, a local handle bearing a globally unique identifier is created on the process on which the constructor was called.
Handles on the other processes are not created immediately.
Instead, we implemented a ``lazy'' allocation mechanism to create the handles of distributed collections on the other processes.
Under this mechanism, the local handle of a collection is allocated on remote hosts the first time a distributed collection is used in an asynchronous activity executed on a remote host.

We resolved these issues by modifying the serialization of our distributed collections such that the table of global ids is checked upon deserialization.
If there are no bindings for the global id of the distributed collection being deserialized, the constructor is called to create the local handle and bind it this global id on this place.
If there was already an object bound to this global id (meaning this is not the first time a closure with this distributed collection is called on this host), then the deserialization resolves to the existing handle.

In the example presented in Listing~\ref{lst:distmapexample}, the local handle for the \verb|dmap| collection on Place~0 is allocated during the construction on line~2.
The handles on the other hosts are created as part of the deserialization of the lambda-expression running from line~4 to~11, prior to its execution on these hosts.

Using this mechanism has the advantage of removing synchronizations over the entire cluster each time a collection is created.
Instead, the local handles of every distributed collection are created little by little as they become necessary.
There is no risk of executing an asynchronous activity on a collection whose local handle is not initialized, as the mere fact that a collection is used in the activity guarantees that the local handle will be created (if it doesn't already exist) as part of this activity deserialization process.

\subsection{Registering entries for relocation}
\label{sec:relocation}

One of the key features of our distributed collections library lies in its ability to relocate entries of a distributed collection between its handles.
Our library builds on and expands a scheme first developed in X10~\cite{collectiverelocation}.

As briefly introduced in Section~\ref{sec:collection}, the \verb|CollectiveMoveManager| can be used to transfer entries belonging to one or multiple collections between all or a subset of the processes participating in the computation, this group being specified at construction using a \verb|TeamedPlaceGroup| instance.
The transfer is initiated when the \verb|sync()| method is called on all the places of the group it operates on.
This call is blocking until it is called on all places involved in the relocation.
As such, the collective relocator mechanism is a synchronization point between asynchronous activities participating in the computation.

The novelty with our library compared to the original scheme lies in the variety of ways programmers can register entries for relocation.
These methods are defined through modular interfaces implemented by our various collections, improving consistency and reducing future development effort.
They allow programmers to specify what entries need to be relocated by specifying relevant arguments and the ``move manager'' used to perform the transfer.

Let us introduce the program of Listing~\ref{lst:rotation} to illustrate the various ways entries of our distributed collections can be marked for relocation.
This program demonstrates a single collective relocation used to relocate objects belonging to multiple collections.
For the sake of simplicity, we chose to make each process send entries to its neighboring $(rank+1)\%n$ process, but this is in no case a limitation of the relocation system as entries originating from a process can be relocated to multiple other processes.

\begin{lstlisting}[float,caption=Rotation of entries between processes using a collective relocator,label={lst:rotation}]
final DistBag<Integer> bag;
final DistChunkedList<Element> cl;
final DistMap<String, String> map;

TeamedPlaceGroup world = 
	TeamedPlaceGroup.getWorld();
final int n = world.size();
world.broadcastFlat(() -> {
  // Prepare the collective relocator
  CollectiveMoveManager mm =
  	new CollectiveMoveManager(world);
  Place destination = place((here().id + 1)%n);

  // Relocation in bulk
  bag.moveAtSyncCount(20, destination, mm);

  // Relocation by range
  for (LongRange range : cl.ranges()) {
    cl.moveRangeAtSync(range, destination, mm);
  }

  // Relocation by key->destination function
  Function<String, Place> relocationRule = 
  	(S9tring key) -> destination;
  map.moveAtSync(relocationRule, mm);

  mm.sync(); // Perform the transfer
});
\end{lstlisting}

\emph{Relocation in bulk} is available to all of our distributed collections. They feature a method called \verb|moveAtSyncCount| which is used to relocate the specified number of entries.
The library decides which entries are relocated without the input of the programmer.
In Listing~\ref{lst:rotation}, this method is used to transfer~20 entries contained in each \verb|bag| handle to their neighbor on line~15.
This is the only available relocation method for the distributed set \verb|DistBag<T>| as individual entries in this collection are devoid from any ``identity''.

\emph{Relocation by range or by key} is possible for distributed collection in which entries are identified by a unique identifier.
We distinguish between collections where entries can be identified by a key, such as \verb|DistMap| and \verb|DistMultiMap|, and collections where entries can be designated through an entire range, such as \verb|DistChunkedList| and its derivatives.
In our library, this is enforced using two generic interfaces \verb|RangeRelocatable<R>| and \verb|KeyRelocatable<K>| which define a number of signatures for methods \verb|moveRangeAtSync| and \verb|moveAtSync| respectively.

We demonstrate the relocation using a range on line~18 of Listing~\ref{lst:rotation}.
Using the loop of lines~18-20, all the ranges contained in collection \verb|cl| are marked for relocation to the neighboring host.
It is not an obligation to specify a range which corresponds exactly to a ``chunk'' contained by the local handle.
Programmers can specify a range which either spans several of the ``chunks'' contained in the local handle or is a sub-range of a single chunk.
In this case, the existing chunks will be split as necessary before relocation.

On line~25, the entries of the distributed map \verb|map| are all marked for relocation using the \verb|relocationRule| function defined just above.
Internally, the \verb|relocationRule| function is applied to each key contained in the local handle to determine their respective destination.
In this example, the ``key'' parameter is not used in \verb|relocationRule| which always return the same \verb|Place| object as the destination, but more sophisticated implementations are entirely possible.

\subsection{Communication patterns for entry relocation}
\label{sec:serializationdeserialization}

When registering some entries for relocation into a \emph{move manager}, our library actually registers a pair of serializer and deserializer into the move manager instance provided as argument.
When the \verb|sync()| method of the collective relocator is called, the serializer is called to convert the targeted objects into bytes. The deserializers are also written to the byte array.

In a collective relocation, each place therefore obtains an array of bytes (possibly empty) to send to every other place participating in the computation.
The transfer of objects is then performed in two steps.
First, the number of bytes to be sent by each process participating in the transfer is exchanged with an MPI \verb|Alltoall| call using the underlying communicator of the \verb|TeamedPlaceGroup| specified with the constructor of the \verb|CollectiveMoveManager|.
This allows each process to know how many total bytes to expect and prepare buffer arrays of the appropriate size.
Then, the byte arrays are exchanged between the processes using an MPI \verb|Alltoallv| call.
Each host then proceeds to deserialize the bytes it received and place the entries into their respective collection handle.
Due to the blocking MPI calls used to perform the relocation, the \verb|sync| method of the \verb|CollectiveMoveManager| is a synchronizing call between asynchronous activities running on different processes.

The same general process is used to implement other features of the library.
In the case of the market replication in PlhamJ shown in Listing~\ref{lst:cachablearray}, the closures provided as argument to the \verb|broadcast| method are used to produce the objects being transferred (in this case instances of class \verb|MarketUpdate|) and to update the \verb|Market| replicas located on the remote host.
Our library takes care of serializing and deserializing the objects used as intermediary vessels.
Then, MPI \verb|Bcast| calls are used instead of \verb|Alltoall| as the order-handling process is the sole source of information.
Similarly, the order relocation performed in Listing~\ref{lst:orderproduction} relocates all the entries of collection \verb|orderBag| to the first process of the distributed program.
After the serialization of the entries to transfer, \verb|Gather| and \verb|Gatherv| MPI calls are used as there is only one ``recipient'' in this communication pattern.

\subsection{Using the library, compilation and execution}
\label{sec:usinglibrary}

Our library comes in the form of a Maven project available on GitHub under the terms of the Eclipse Public License v1.0 at the following url: \url{https://github.com/handist/collections.git}.
At the time of writing, the current version is \emph{v1.2.0}.
All the necessary dependencies used by our project, namely a slightly customized version of the APGAS for Java library and the MPJ-Express library, are downloaded automatically. 
It is therefore sufficient to add our library as a dependency to any Java project to be able to compile programs that use our library, including on systems on which MPI is not installed.

As we rely on the MPJ-Express library~\cite{mpjexpress} to provide the MPI calls to our program, it is necessary for the ``native'' part of this library to be compiled beforehand on the execution environment.
Fortunately, this is thoroughly explained in the MPJ-Express documentation.

Listing~\ref{lst:launchprogram} shows a generic command used to launch a program with our library.
Programs are launched with the \verb|mpirun| command as can be seen on line~1.
The number of processes and their allocation on hosts are specified with the usual MPI options.
In the example shown on Listing~\ref{lst:launchprogram}, 4 processes allocated according to the specified hostfile are used.

The Java command is then used to launch the processes part of the computation.
The classpath is specified as per usual using the \verb|-cp| option on line~2. 
On line~3, the location of the MPJ-Express shared library is specified using the \verb|-Djava.library.path| option.
As per the MPJ-Express compilation instructions, this shared library is customarily placed under the \verb|${MPJ_HOME}/lib| directory. 

We provide a specific launcher with our library which takes up the role of the main class, as can be seen on line~4.
The user's main class is then passed as the first argument to our launcher, with the programs arguments following after that.

\begin{lstlisting}[float,caption={Command used to launch a program with our distributed collection library}, label={lst:launchprogram}]
mpirun -np 4 --hostfile ${HOSTFILE} \
	java -cp collections-v1.2.0.jar:program.jar \
	-Djava.library.path=${MPJ_HOME}/lib \
	handist.collection.launcher.Launcher \
	${MAIN_CLASS} ${ARG1} ${ARG2}
\end{lstlisting}

\section{Evaluation}
\label{sec:evaluation}

The goal of our evaluation is threefold.
First, we want to establish the greater programmability of our library when writing distributed programs.
Secondly, we want to establish the performance of programs written with our library against equivalent ones.
Lastly, we want to verify that load-balancing techniques made possible by our library are capable of adapting distributions to match uneven or evolving cluster performance.

We use three applications\footnote{All our programs are freely available on GitHub in the following repositories: 
	\begin{itemize}
		\item	\url{https://github.com/handist/collections-benchmarks}
		\item	\url{https://github.com/plham/plhamJ}
	\end{itemize}
} for the purposes of our evaluation, the K-Means benchmark adapted from the Java Renaissance benchmark suite~\cite{renaissance}, the N-Body molecular simulation \emph{MolDyn} adapted from the Java Grande benchmark suite~\cite{javagrande}, and our financial market simulator \emph{PlhamJ}. All three were presented in Section~\ref{sec:motivatingcases}.

We first discuss matters related to programmability in Section~\ref{sec:programmability}. We then compare the performance of the original K-Means and MolDyn implementation against the versions we implemented with our library in Section~\ref{sec:performance}. Finally, we establish the capabilities of our high-level load-balancing features using PlhamJ in Section~\ref{sec:loadbalancing}.
We used OACIS~\cite{oacis} to manage the large number of executions necessary for this evaluation.

\subsection{Programmability}
\label{sec:programmability}

Programmability is a difficult criteria to judge.
Comparing programs using quantitative criteria such as \emph{lines of code} (loc) can be done, but such criteria alone cannot be used to determine whether some model or library is beneficial or not.
An abstraction supporting a particular pattern may reduce the amount of code necessary, but if it is too specific or convoluted to be used in other applications the claim of better programmability is weak. 
On the other hand, qualitative criteria may be controversial or subject to a certain level of subjectivity. 

We believe our library brings significant gains in programmability thanks to three key characteristics: (1) its support for local parallelism, (2) the notion of ``teamed operation," and (3) the high-level support for distribution management.
Concerning the support for local parallelism, as we demonstrated in Section~\ref{sec:motivatingcases}, our distributed collections provide multiple parallel methods taking closures as arguments.
This approach re-centers programs on the actual computation at hand rather than how the parallelism is supported.
On this matter, the comparison between the K-Means implementation with our library and the original Java Renaissance suite~\cite{renaissance} is particularly interesting. 

In the Renaissance K-Means implementation, the points are managed by range explicitly in order to implement the ``recursive task'' implementation required by the Java \verb|ForkJoinPool|.
By comparison, the range management remains totally internal to the \verb|ChunkedList| class we use in our implementation. 
The range management remains also absent from the classes used to support the reductions needed by the algorithm. 
As a result, the total size of the distributed K-Means written with our library (excluding argument parsing and initialization) amounts to just over 200 lines of code compared to over 400 lines of code for the Renaissance implementation.
Moreover, the legibility of the program is entirely preserved despite its distributed nature, as made evident by Listing~\ref{lst:kmeans}.

Moreover, the management by the library of thread-dedicated data structures greatly simplifies programs for what would otherwise become a cumbersome implementation.
Our library allocates just the necessary data structures to support the number of available threads on the system.
This remains entirely transparent to the programmer who may use the various parallel methods knowing that the appropriate number of threads will be spawned even if the number of threads available varies from a host to another. 

The second gain brought by our library comes by the introduction of teamed operations on our distributed collections.
These methods define the scope of their intervention by using either the group of processes on which the supporting collection is defined, or by specifying the group explicitly in a constructor (as is the case for the collective relocator). 
This contributes to the clear identification of both the hosts that are involved in said teamed operation and the synchronizing point between the asynchronous activities running on different host.

This is particularly evident in the case of PlhamJ, where the first hosts performs different tasks than the others.
In this application, the teamed methods both serve as the necessary communication support to implement the program but also as the synchronization point used to determine completion of a remote procedure.
For instance, the order submission of agents cannot start until the teamed market information broadcast is received by the local host. 
Similarly, the order-handling on the first process cannot start until the orders submitted by each agent for this round are received through the teamed gather operation.

Finally, programmers have complete and dynamic control over the entry distribution of the distributed collections.
The high-level relocation abstractions we provide makes this management easy, with supporting features such as the distribution tracking countering the challenging nature of a dynamic distribution where necessary.
The most prominent example of this lies in PlhamJ where agents are relocated from hosts to hosts to balance the computational load while the distribution tracking ensures that information meant for a specific agent reaches its destination.
Internally, the management of entries by range makes this both elegant and efficient.

\subsection{Performance comparison against original benchmark implementation}
\label{sec:performance}

To verify that our distributed collections library provides reasonable performance, we compare the performance of two programs written with our library against reference benchmark implementations of K-Means~\cite{renaissance} and MolDyn~\cite{javagrande}. 
We conduct our performance evaluation on the OakForest-PACS supercomputer which features 68 core Xeon Phi CPU, using up to 64 compute nodes.
The hardware and software environment used on the OakForest-PACS supercomputer are summarized in Table~\ref{tbl:ofp}.
\begin{table}
\caption{Hardware and Software environment on the OakForest-PACS supercomputer}
\label{tbl:ofp}
\begin{tabular}{lp{6cm}}
\hline
Property & Value \\
\hline
Processor & Intel Xeon Phi 7250 (1.4 GHz, 68 cores) \\
RAM & 96GB DDR4 \\
Java version & Open JDK 1.8.0\_222 \\
MPI version & Intel MPI with MPJ-Express v0\_44 Java native bindings \\
\end{tabular}
\end{table}

\subsubsection{K-Means}

\begin{table}
	\caption{KMeans benchmark parameters}
	\label{tbl:kmeansparameters}
	\begin{tabular}{lcc}
		\hline
		Configuration & ``small'' & ``large'' \\
		\hline
		Nb of points/host & \multicolumn{2}{c}{10 million} \\
		Point dimension & 3 & 5 \\
		Number of clusters & 50 & 2000 \\
		Iterations & \multicolumn{2}{c}{30} \\
	\end{tabular}
\end{table}

The original Renaissance benchmark operates on a single process.
We compare it against two implementations of K-Means prepared with our library: a ``single-host'' version, and the distributed ``teamed'' version previously introduced in Section~\ref{sec:motivatingcases}. 

We perform our evaluation in weak scaling from 1 to 64 hosts, increasing the number of points proportionally to the number of hosts involved in the computation.
We run the K-Means algorithm for 30 iterations and compare the iteration time between the implementations.
The details of the program parameters we used are shown in Table~\ref{tbl:kmeansparameters}.

\begin{figure}
\begin{subfigure}[h]{\columnwidth}
  \includegraphics[height=\linewidth,angle=270]{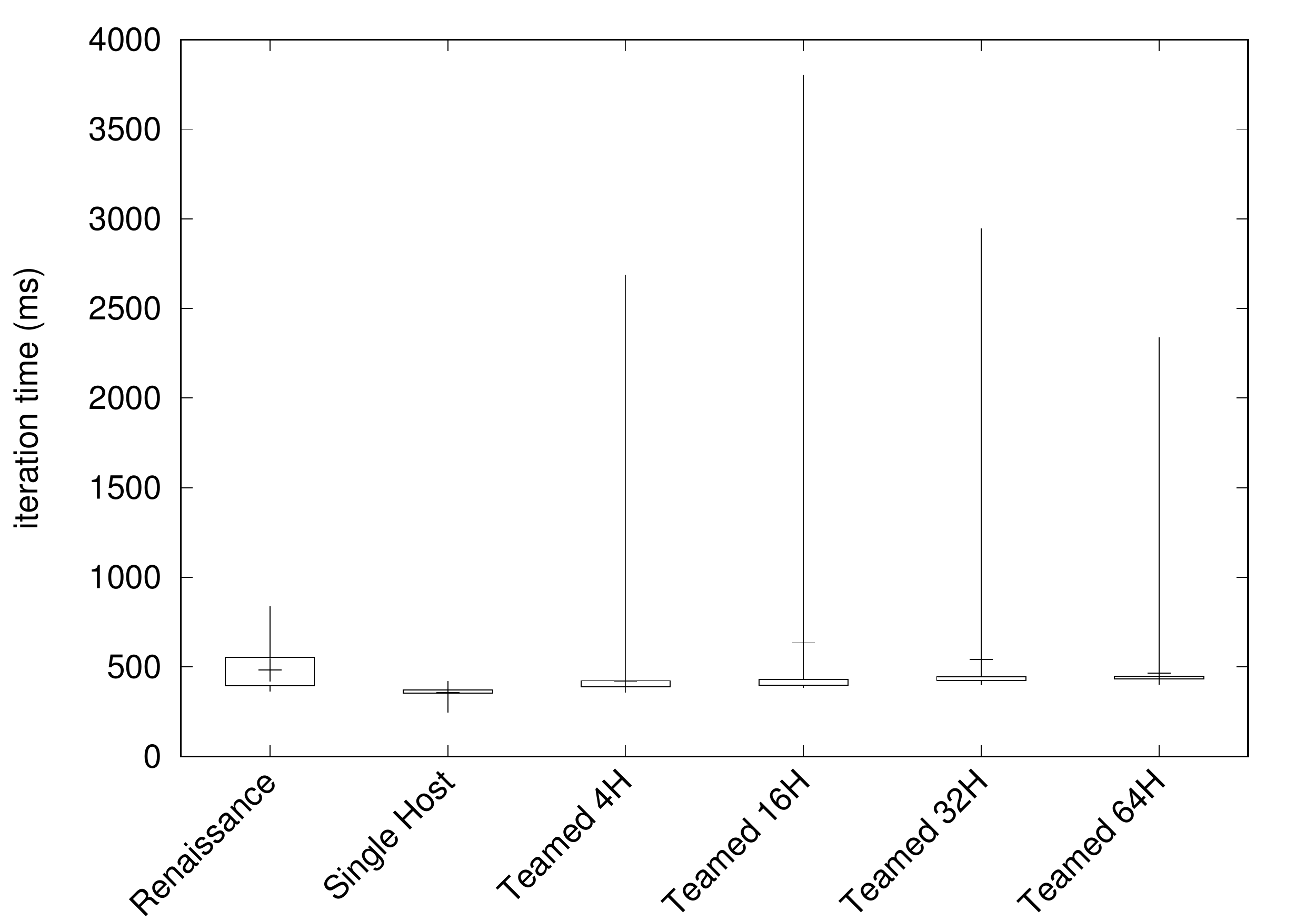}
  \subcaption{small configuration}
  \label{fig:small-kmeans}
\end{subfigure}
\begin{subfigure}[h]{\columnwidth}
  \includegraphics[height=\linewidth,angle=270]{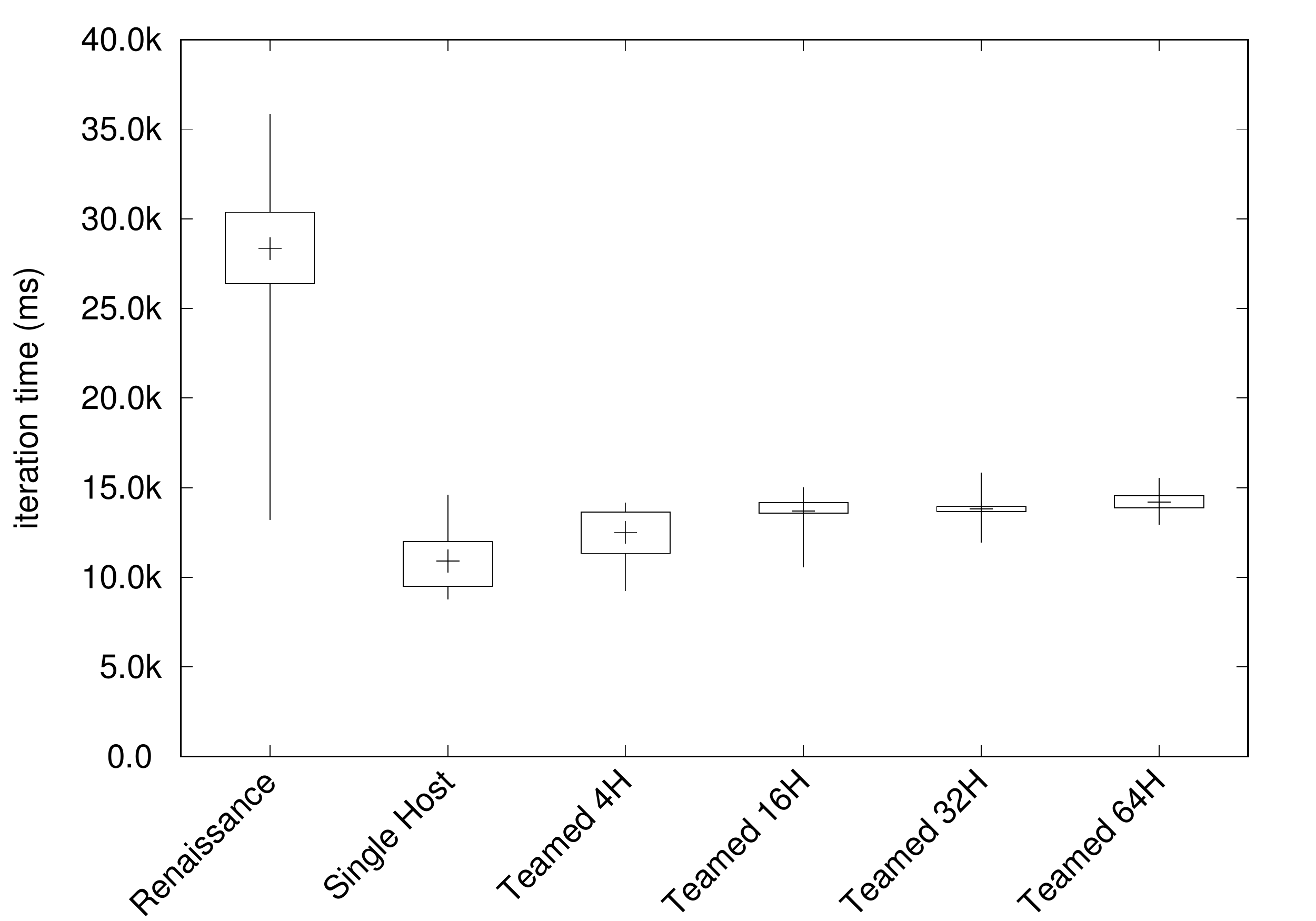}
  \subcaption{large configuration}
  \label{fig:large-kmeans}
\end{subfigure}
\caption{KMeans iteration times. The brackets and boxes represent the minimum, 1st quartile, 3rd quartile, and maximum values while the cross corresponds to the average value of 5 sample runs or 30 iterations each.}
\label{fig:kmeans}
\Description{}
\end{figure}

The results are presented in Figure~\ref{fig:kmeans} where we plot the minumum, first quartile, third quartile, maximum, and average iteration time obtained with each program version.

With the "small" parameter configuration, the average iteration time is kept just below 500ms with the Renaissance benchmark.
Our implementation on a single host is 20\% faster. 
This higher performance is maintained on 4 hosts (12\% faster than Renaissance) despite the communication needed by the reductions. 
However, our implementation is not capable to scale further with such short iteration times.
Over the course of the ``teamed'' program executions we witnessed a few particularly long iterations, the longest of which occurred on a 16 host execution and lasted just under 4 seconds.
This drives the average iteration time upwards despite the overwhelming majority of iterations completing within 500ms. 

We are not certain as to what causes this phenomenon. 
We believe it could be explained by some of the processes in the cluster performing garbage collection with unfortunate timing and delaying the communication with the other processes during the teamed reductions of the program.
This would in turn delay the progress of the entire program as the other processes are stuck waiting on the result of the reduction.

In the second "large" K-Means, the number of clusters is dramatically increased compared to the "small" parameter.
As a result the computational load consisting of assigning each point to a cluster becomes predominant over the two reductions and the iteration times increase for the Renaissance version to an average of 28.3 seconds.
Our single host implementation however, maintains an average iteration time far below, just under 11 seconds.

This performance gap between the Renaissance implementation and our Single Host implementation can be explained by the higher memory consumption (and more frequent garbage collection) of the Fork/Join implementation.
While the Renaissance version is capable of delivering short iteration times, as is made clear by the minimum iteration time of 13.2s, it is not capable of sustaining them over the entire course of the execution.
Our distributed implementation also shows better performance than the original implementation, with the average iteration times kept below 15s up to 64 hosts.
Under this higher computational load, the performance trouble witnessed under the ``small'' parameter configuration is absorbed, with the iteration times of obtained in the 64 hosts configuration only 30\% longer than our ``single host'' version, but still half that of the Renaissance implementation.

\subsubsection{MolDyn}

The original Java Grande version of MolDyn built on MPI uses 1 thread per host.
Against the original version we compare two versions implemented with our library: a single-threaded version (Handist ST) similar to the original implementation, and a hybrid version (Handist Hybrid) which uses multiple threads on each process.

\begin{figure}
  \includegraphics[height=\columnwidth,angle=270]{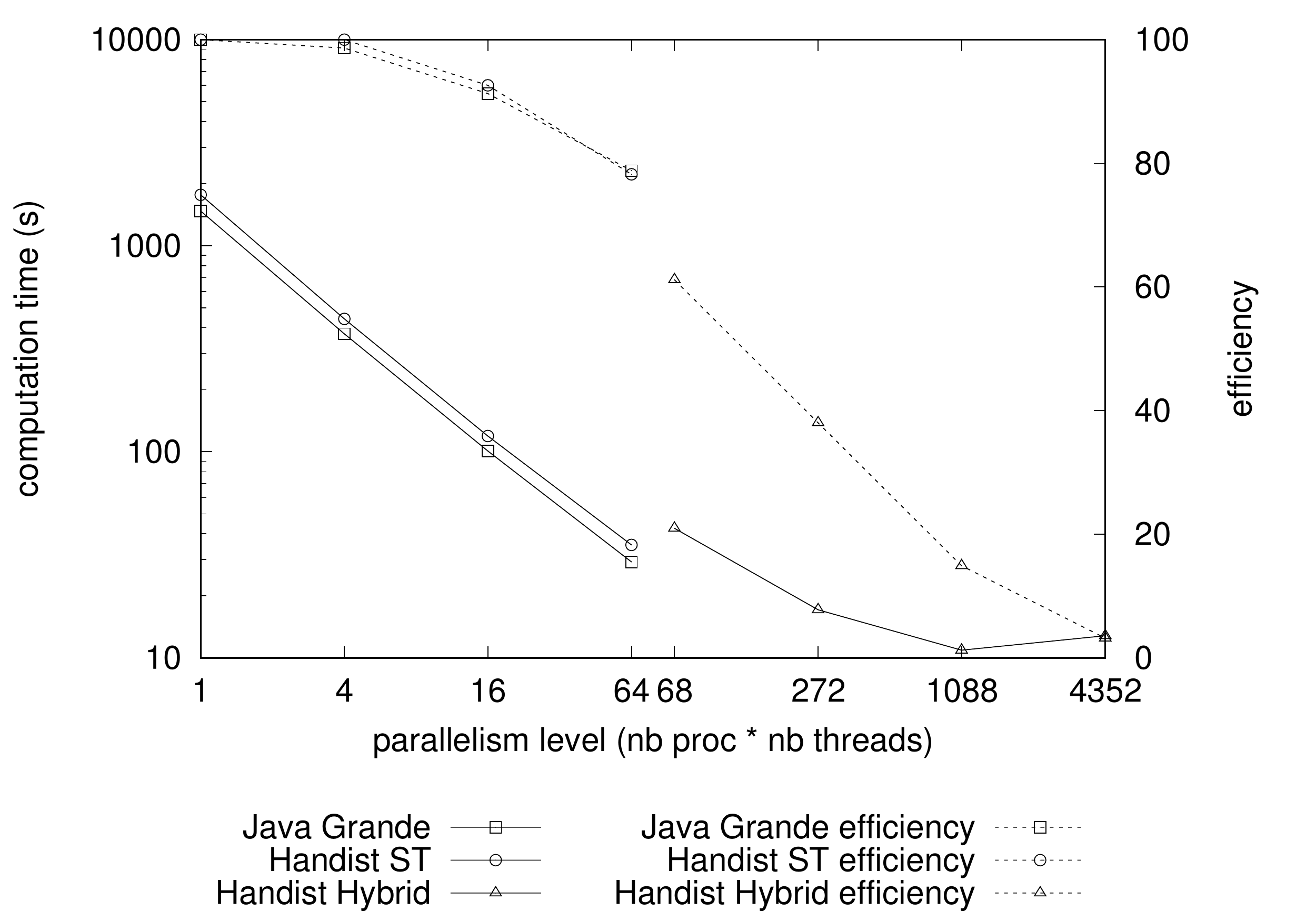}
  \caption{Computation time and efficiency of the MolDyn benchmark on the OakForest-PACS supercomputer}
  \label{fig:moldyn}
  \Description{}
\end{figure}

We run the MolDyn benchmark in strong scaling (same problem size for increasing cluster size) on the OakForest-PACS supercomputer from 1 to 64 hosts with 32,000 particles.
We use 68 threads per process for our hybrid implementation, resulting in its parallelism level with a single process to be slightly higher than the Java Grande and the ST version with 64 hosts.
We measure the total computation time of the simulation after a short warmup.
The computation times and the efficiency of each program version are are presented in Figure~\ref{fig:moldyn}.
An ideal efficiency of 100\%, i.e. perfect scaling, would mean that increasing the computational resources by a factor $n$ yields execution times $n$ times shorter.

\begin{figure}
\includegraphics[height=\columnwidth,angle=270]{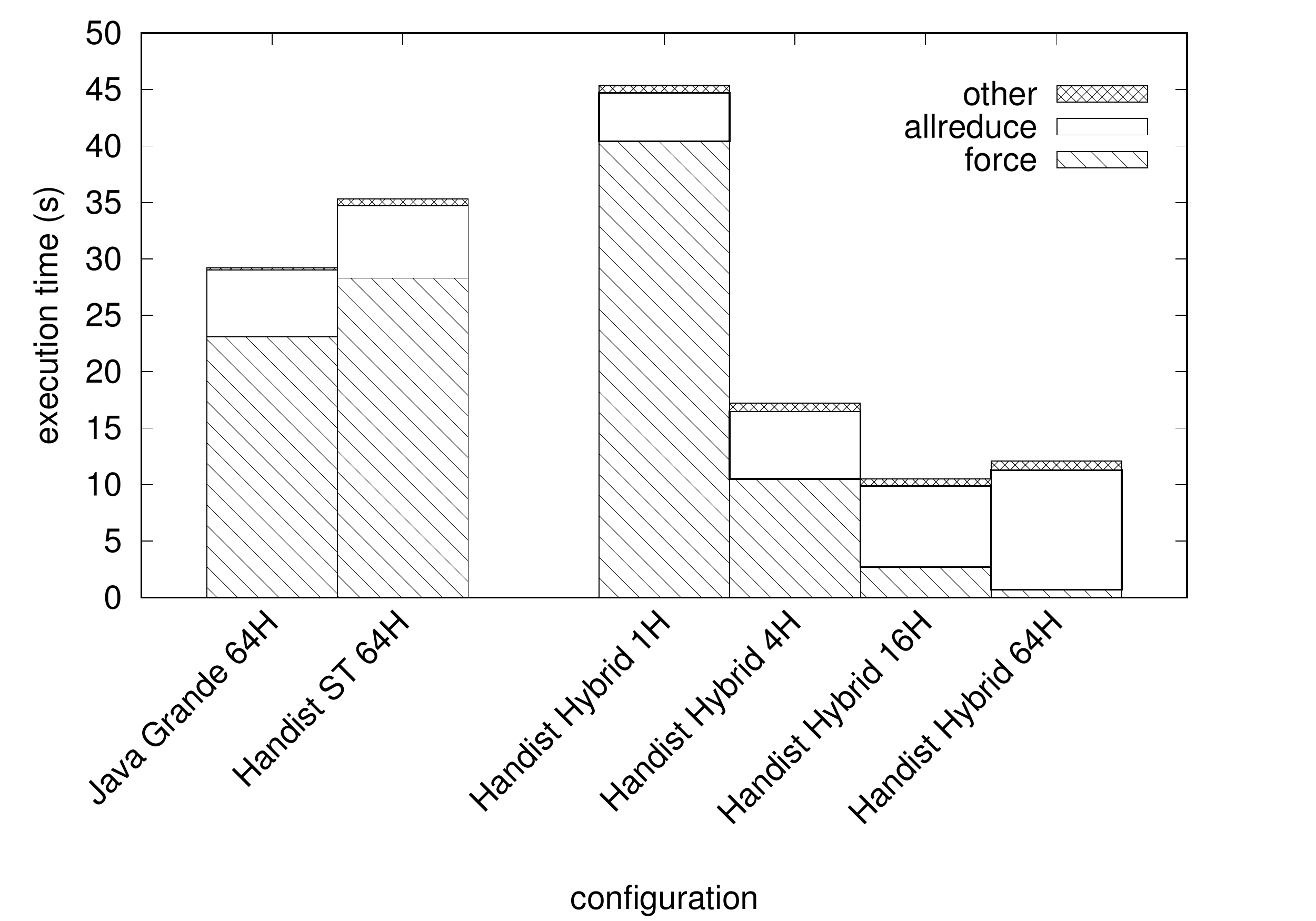}
\caption{Computation time breakdown of the higher-parallelism executions of the MolDyn benchmark}
\label{fig:moldynbreakdown}
\Description{}
\end{figure}

First, comparing the Java Grande version against our single-thread (ST) implementation, we note a 20\% increase in computation time.
We believe this is a reasonable amount of overhead considering the fact we moved away from the primitive type arrays to use objects to store the particles in our ST and hybrid implementations.
The efficiency for both the ST and Java Grande versions follow the same pattern, decreasing down to 78\% on 64 hosts.

This can be explained by the nature of the computation at hand. 
In all three versions of MolDyn studied here, the time taken by the ``allreduce'' sum of the forces across hosts takes a total of about 5 seconds of the total computation time, irrespective of the number of hosts or threads used.
On the Java Grande and Handist ST executions from 1 to 16 hosts, the computation time was dominated by the force computation. 
As can be seen in Figure~\ref{fig:moldynbreakdown}, this is no longer the case on 64 hosts where the ``allreduce'' part represents about 15\% of the computation time.
As the parallelism increases and the force interaction computation time decreases, this incompressible part of the program takes up a relatively larger part of the total computation time, decreasing efficiency.

Secondly, our hybrid implementation shows a slightly different efficiency pattern compared to the other implementations. 
Its efficiency for the 1 host/68 threads configuration loses an additional 17 percentage points of efficiency compared to the similar level of parallelism of the ST version running on 64 hosts.
This is mostly imputable to the overhead brought about by the use of the accumulator mechanism in the hybrid version. 
Also, the fact that we used an entire host for each single-threaded Java Grande and Handist ST gives those versions a certain advantage.
In future work, we hope to be able to reduce the overhead brought by the use of the accumulator mechanism by introducing alternative implementations that would only allocate ranges on a per-need basis rather than allocating the complete range from the start.

We are able to further reduce the execution time down to just over 10 seconds with the hybrid version running on 16 hosts (1088 total threads), albeit with decreasing efficiency.
The execution on 64 hosts shows it is counterproductive to stretch the program any further, with the total computation time increasing from 10 to 12 seconds. 
As can be seen in Figure~\ref{fig:moldynbreakdown}, the computation time is dominated by the ``allreduce'' part of the computation on hybrid executions with larger parallelism.

\subsection{Dynamic Load Balancing in PlhamJ}
\label{sec:loadbalancing}

The objectives of the evaluation conducted with our PlhamJ distributed financial simulator is twofold. 
First, we want to demonstrate the capability of a distributed program to adapt itself to the uneven performance of the cluster on which it is runs thanks to the features of our library.
Second, we want to verify that the load-balancing measures we implemented in PlhamJ are able to react to dynamic changes in the cluster performance.

We perform the evaluation on our Beowulf cluster composed of two types of hosts: ``piccolo'' hosts which feature a 4-core CPU, and the higher-parallelism ``harp'' host which features two 12-core CPUs.
The detailed hardware characteristics are outline in Table~\ref{tbl:beowulf}.
We use up to 5 hosts in three different cluster configurations summarized in Table~\ref{tbl:clusterconfigs}.

In \emph{Config~A}, we use a typical approach consisting of allocating one process oneach ``piccolo''.
The order-processing process is allocated on one host, while the three other hosts are dedicated to agents' order submission.

In \emph{Config~B}, we allocate an additional agent-processing process on the host holding the order-processing host (5 processes on 4 piccolos).
This choice of allocation can be justified by the fact that the process which handles the orders remains idle while the agents are making their submission.
There is therefore some amount of computing resources left untapped on the server hosting the handling of orders in which we try to leverage with this second configuration.

Finally, in \emph{Config~C}, we add the ``harp'' host as an order-handling process compared to Config~B.
The challenge of Config~C lies in the nature of this additional server which brings more parallelism than the identical ``piccolo'' hosts used so far. 
It is therefore difficult to predict a priori what a good distribution of agents should be with such a cluster configuration.

To simulate dynamic changes in performance, we introduced a parasite program called ``Disturb''.
This program runs concurrently to our simulator and computes an artificial 20 seconds load on one of the hosts.
When the 20 seconds have elapsed, another host is chosen as the victim.
The sequence of hosts ``disturbed'' by this program is deterministic following an initial seed to allow us to reproduce its effects over multiple executions.

We compare the performance of our ``level extremes'' load-balancing strategy previously discussed in Section~\ref{sec:periodiclb} against the fixed uniform distribution without load balance ``no lb''.
The results are presented in Figure~\ref{fig:plhamiteration}.

\begin{table}
\caption{Hardware characteristics of our uneven Beowulf cluster}
\label{tbl:beowulf}
\begin{tabular}{lp{2.8cm}p{2.8cm}}
\hline
Machine Type & ``piccolo'' & ``harp'' \\
\hline
Nb of servers & 4 & 1 \\
Processor & Intel Xeon E3-1230 V2 (3.3GHz, 4 cores) & dual Intel Xeon E5-2680 V3 (2.5GHz, 24 cores combined) \\
RAM & 16GB DDR4 & 128GB DDR4 \\
\hline
Java version & \multicolumn{2}{c}{OpenJDK v1.8.0\_312} \\
MPI version & \multicolumn{2}{p{6cm}}{Open MPI v3.1.6 with MPJ-Express v0\_44 Java native bindings} \\
\end{tabular}
\end{table}

\begin{table}
\caption{Cluster configuration summary}
\label{tbl:clusterconfigs}
\begin{tabular}{cp{5.5cm}}
\hline
Configuration & Description \\
\hline
Config A & 4 processes on 4 piccolos (no load unbalance expected) \\
Config B & 5 processes on 4 piccolos (piccolo 0 hosts the order-handling process and one agent-handling process) \\
Config C & 6 processes on 4 piccolos and 1 harp (piccolo 0 hosts the order-handling process and one agent-handling process) \\
\end{tabular}
\end{table}

\begin{figure}
\includegraphics[width=\linewidth,trim={0 10 0 0},clip]{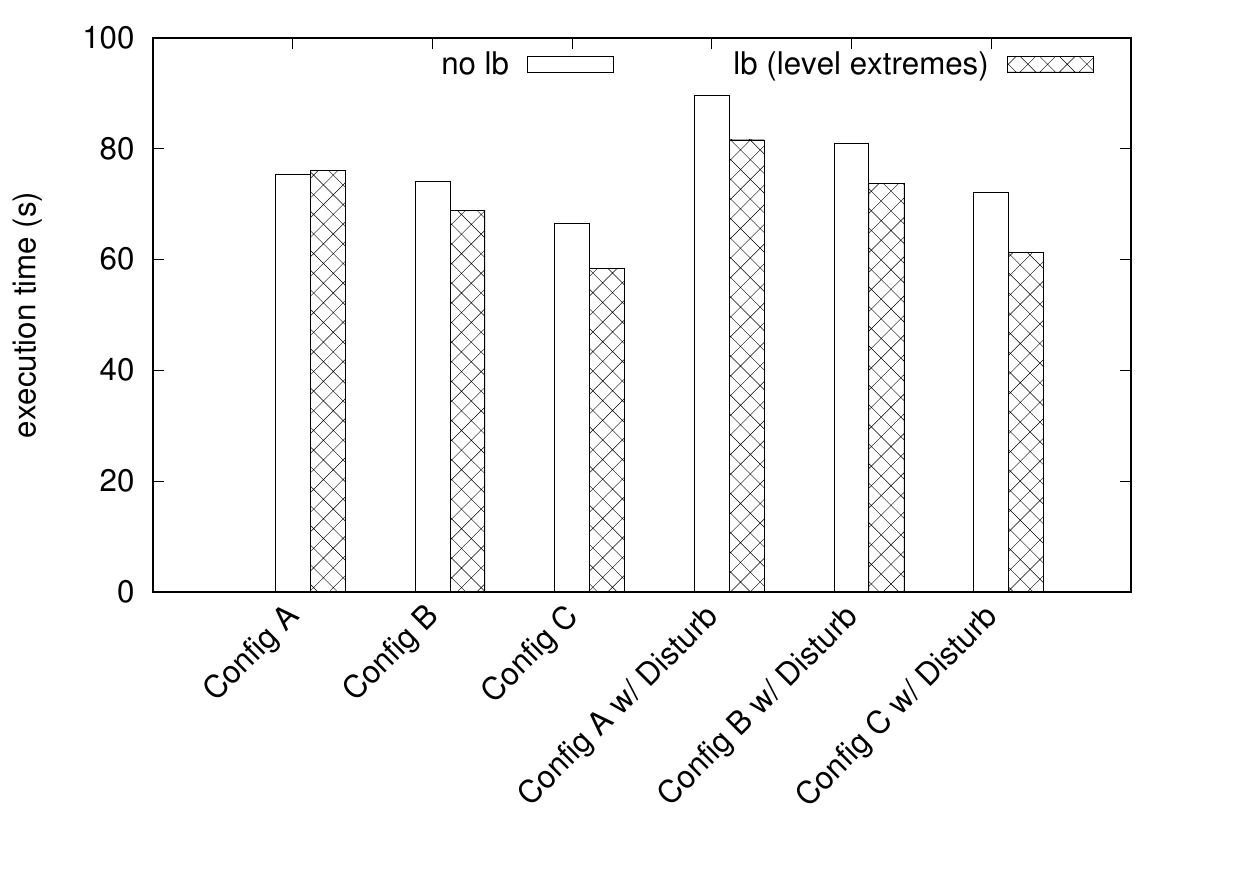}
\caption{Execution time of the Plham simulation depending on the cluster configuration}
\label{fig:plhamclusters}
\Description{
This figure is an histogram comparing the execution time in seconds of the ``no load balance'' and ``levelextremes'' load balance strategies (or lack thereof) on the 3 cluster configurations ConfigA, ConfigB, and ConfigC, with and without the competing ``disturb'' program.
We summarize in a table format the same information below:

"Configuration"  "no load balance"  "level extremes"
"configA"  75.35  75.97
"configB"  74.04  68.88
"configC"  66.49  58.38
"configA w/ disturb"  89.66  81.52
"configB w/ disturb"  80.96  73.74
"configC w/ disturb"  72.01  61.30
}
\end{figure}

We are able to draw two conclusions from the PlhamJ executions without the \verb|Disturb| program.
First, our basic load-balancing incurs no overhead in our distributed PlhamJ simulator as demonstrated under the ``Config A'' results.
Execution times for the static and the load-balanced version are almost identical at 75.3 and 76.0 seconds in this configuration where no load-balancing is required.
This can be explained by the fact that the (hypothetical) transfer of Agents between hosts takes place concurrently to the order-handling on the first process.
In our experience, the transfer of Agents completes before the order-handling and thus does not negatively impact performance.

Secondly, this basic load-balancing technique is capable of handling an uneven cluster configuration, as can be seen in the execution time of PlhamJ under Config~B and Config~C. 
The ``level-extremes'' strategy perform better than its counterpart with its computation time shorter by 8 and 12\%.

\begin{figure}
\begin{subfigure}[h]{\columnwidth}
\includegraphics[width=\linewidth]{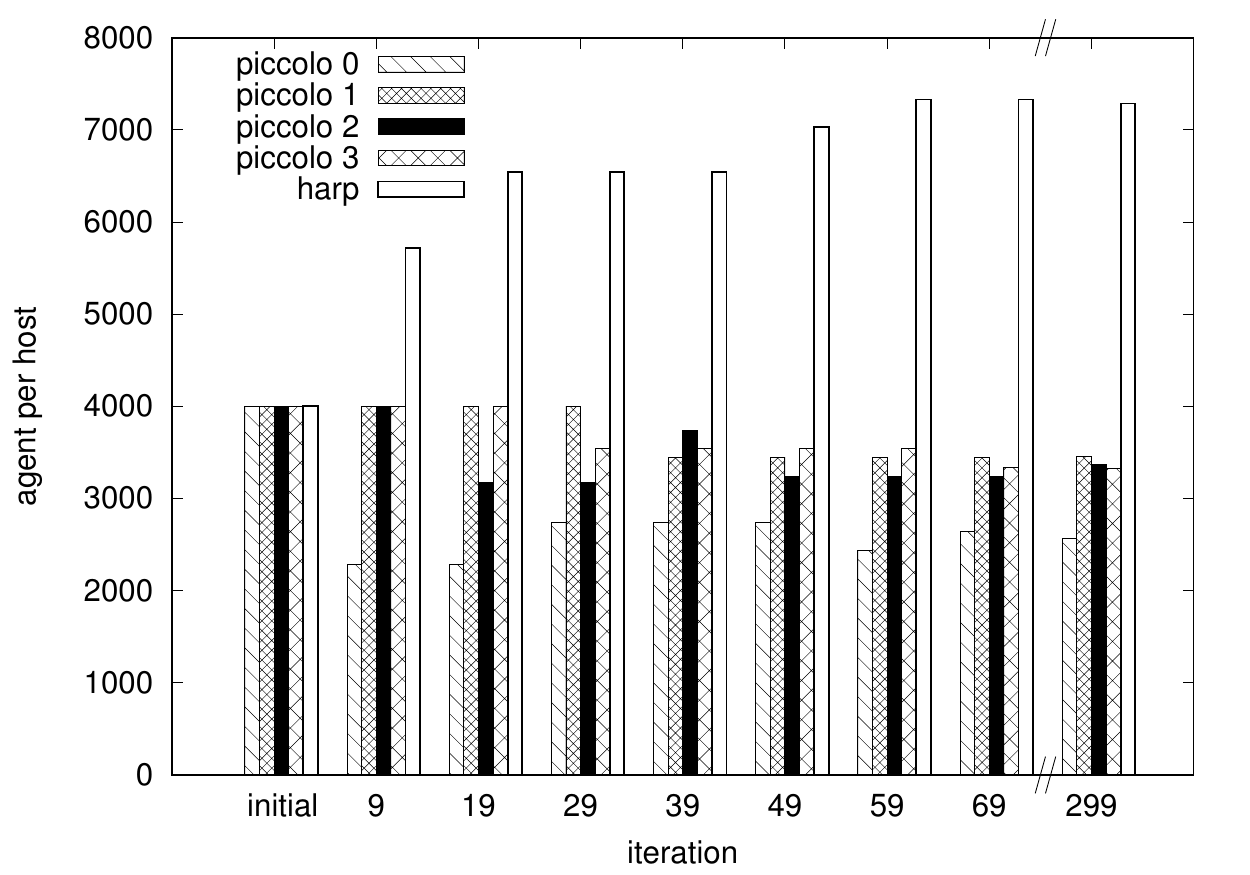}
\caption{under Config C cluster configuration w/o disturb}
\label{fig:configc}
\Description{}
\end{subfigure}
\begin{subfigure}[h]{\columnwidth}
\includegraphics[width=\linewidth]{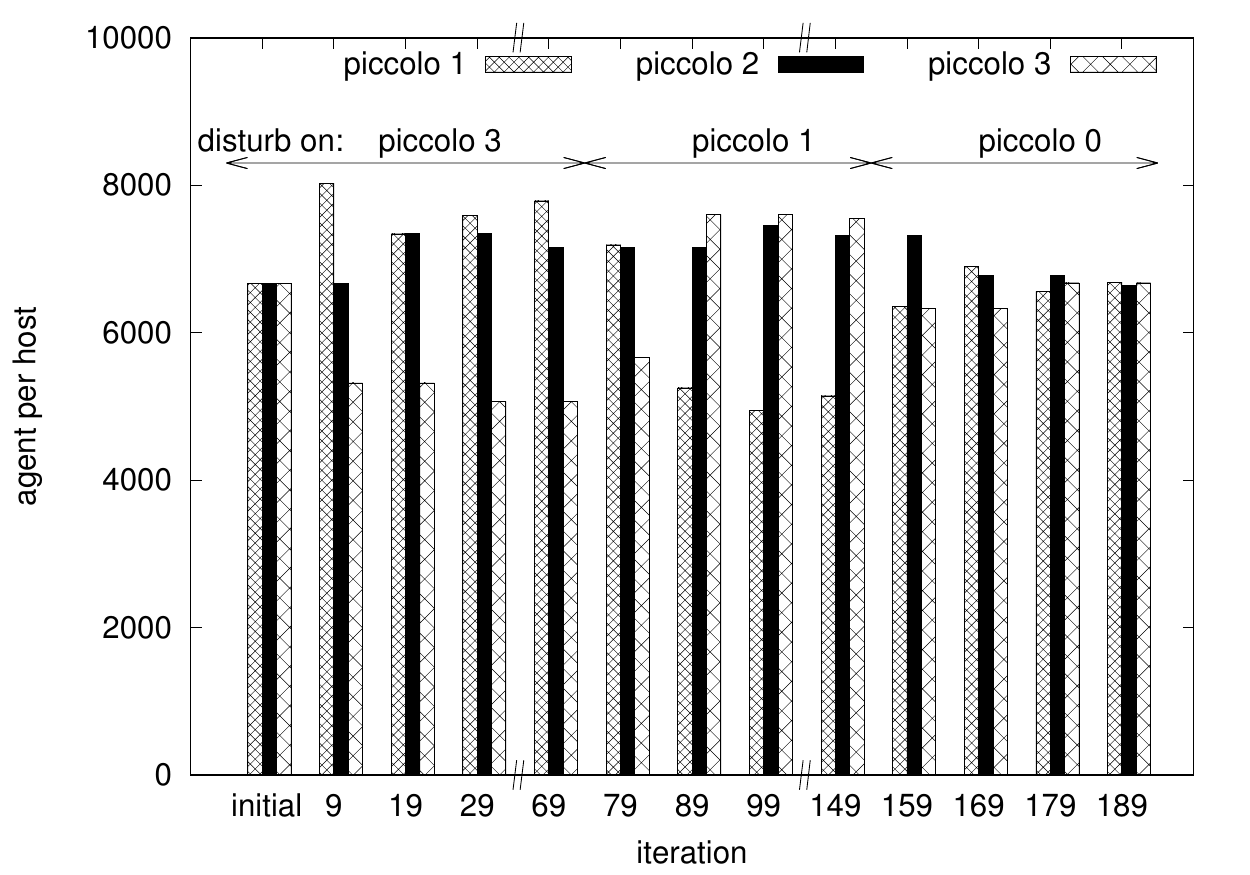}
\caption{under Config A cluster configuration w/ disturb}
\label{fig:configawithdisturb}
\end{subfigure}
\caption{Distribution of agents over time}
\label{fig:distribevolution}
\Description{}
\end{figure}

Depending on the configuration, our load balancing strategy delivers execution times between 7 and 15\% shorter than the fixed uniform agent distribution.
The distribution of agents over time during an execution under Config~C is presented in Figure~\ref{fig:configc}.
The distribution becomes stable after only 30 iterations (4 seconds into the simualtion).
Seeing as piccolo~0 hosts both the order-handling process and an agent-handling process, it ends up containing fewer agents than its piccolo~1-3 counterparts.
Also, the higher parallelism available to the process allocated on our ``harp'' server is made evident by the fact it obtains over a third of the total agents in the simulation.

The experiments with the parasite program presented in Figure~\ref{fig:plhamclusters} also show that our basic load balancing strategy is capable of handling dynamic changes in the cluster performance, with execution times between 8 and 15\% shorter depending on the configuration.

In Figure~\ref{fig:configawithdisturb}, we show the evolution of the agent distribution under Config~A w/ Disturb. Under this configuration, the only source of disparities between the hosts performance is the presence of the parasite program on one of the hosts.
At the beginning of the simulation, the server hosting process piccolo~3 is being disturbed, resulting in some of its agents to be offloaded to the other processes.
Then, starting between the 70th and 80th iteration of the simulation, the disturb process moves to piccolo~1.
As a result, agent are moved away from piccolo~1 and the previously disturbed piccolo~3 is assigned more agents.
In the last part shown on this graph, the disturb program moves to piccolo~0 which hosts the process dedicated to processing the orders.
As a result, there is no longer a discrepancy in the available processing power between the piccolo~1, 2, and 3. 
Our load balancer therefore redistributes the agents evenly between hosts starting from the 160th iteration of the simulation.

\section{Related Work}
\label{sec:relatedwork}

The concept of distributed collections is not new. The work we present here bears resemblance with earlier work from Lee \& Gannon~\cite{pcpp} in which they define the \emph{Distributed Collection Model} for the pC++ programming language.
Under this model, a distributed collection contains elements that can be referenced through a globally unique handle.
A distribution describes how the elements are assigned to the virtual processors used at runtime. Parallelism is supported by sending a message to the collection which will in turn invokes the specified method on all elements of the collection.
It is also possible to send such a message to a subset of the virtual processors.
One peculiarity of this model is the capability for individual elements to obtain information from the structure of the collection (i.e. their position in a 1D array or 2D grid).
One limitation of this programming model is that there is a single main control thread for the program resulting in calls on an entire collection to be synchronous.
Under the APGAS programming model this constraint is relaxed, with the progress of asynchronous activities on various hosts being only halted if some communication between hosts is needed as part of the activity.
While multiple distribution strategies are available in this language, there is also no obvious mechanism that would allow to modify the distribution of a collection.

Charm++~\cite{charmpp,charmppcompared} relies on problem over-decomposition into many ``Chares'' to dynamically relocate them on processing elements based on information obtained through profiling and selectable policies.
This means that the programmer does not have to manage distribution or locality as the control is surrendered to the Charm++ runtime.
While this is certainly acceptable for some applications, others will benefit from the explicit data placement and careful local parallelism that our library provides.
The Charm++ ``NodeGroup'' concept could be used to represent the ``local handle'' of a distributed collection as we introduced it in this article.
The syntax used in Charm++ to specify on which chare some action is performed finds an equivalent in the \verb|asyncAt| construct of the APGAS programming model.
However in Charm++ a \emph{branched chare} needs to be defined on all processors participating in the distributed computation.
There is no support for sending a message to a subset of the processors as this is fundamentally not compatible with the Charm++ programming model which remains agnostic to chare location.
One advantage of APGAS over Charm++ is that the completion of certain asynchronous activities can be elegantly controlled through the finish/async model.
This is important for simulations where a higher level of control over the completion of asynchronous tasks is needed.

Some of the benchmarks we used to demonstrate the performance of our library could be programmed using the Map-Reduce model of Hadoop.
As its core, Hadoop involves over-decomposing a problem in a set of independent tasks which can then be scheduled on a computation cluster.
Some work has shown that Habanero-Java combined with Hadoop can be more efficient both in terms of memory consumption and execution time by taking advantage of multithreading~\cite{multicorehadoop}.
However, the target for our parallel \& distributed collection is different. We focus on a more fine-grained level of parallelism than Hadoop, with programs that present more intractable communication patterns.

Chapel is a programming language developed as part of the DARPA's high productivity computing systems program~\cite{chapel,chapel2}.
It allows distribution of arrays through \emph{Block}, \emph{Cyclic}, and \emph{Cyclic Block} distributions.
With an initial array defined, pieces of it can be relocated using these pre-defined distributions.
However, Chapel does not support this features for maps (or \emph{associative domains} per the Chapel idiom).
We support several variants of distributed maps in our collection library, including a multi-maps, with the capabilities of freely relocating mappings on any host over the course of the program execution.
Deitz et al.~\cite{reducandscans} explored improving the programability and the performance of distributed scans and reductions in Chapel and MPI.
In particular, they supplement MPI with a set of preprocessor directives that automatically generate the code to make user-defined parallel and distributed reductions.

More recently, XscalableMP (XMP)~\cite{xmp} has introduced compiler directives for C and Fortran that allows a program to be distributed and parallelized automatically.
However, the XMP only support distributed arrays where we also support other data structures.
An interesting feature XMP supports is the notion of ``shadowing''.
Given a nested for loop, if the computation needs to access neighboring data the compiler directives of XMP are capable of generating code to access data points which may be located on remote hosts.
We can work around this limitation with our library using ``owner/replica'' schemes, but not in a manner quite as elegant as XMP.

UPC is an extension of the C programming language implementing the PGAS programming model~\cite{upc},
using ``private'' and ``shared'' pointers to denote local and remote data.
UPC's distributed arrays make it easy to spread data across processes in cyclic distributions.
If data accessed through a shared pointer is located on a remote process, the UPC compiler inserts the code necessary to transfer the data, providing the illusion of a shared-memory environment to the programmer.
This can be a source of performance issues, with work focusing on optimizing these communication patterns~\cite{upc2}.

There are a number of parallels to be drawn between UPC++~\cite{upcpp} and our work.
UPC++ is also a PGAS language which provides Remote Procedure Call (RPC) using futures and promises, analogous to the \verb|asyncAt| method used in APGAS for Java.
Unlike UPC from which it is derived, shared pointers cannot be dereferenced directly in UPC++ v1.0, making communication between processes explicit in the program.
The same approach is taken in APGAS for Java where the constructs provided need to be used to access remote memory.
While no abstractions as elegant as the \verb|finish|/\verb|async| is provided to detect quiescence, the use of futures allows programmers to describe which task or computation should be performed after completion of some previous one.
This is lacking in the system we use, with the \verb|finish| construct most useful in cases where recursive/transitive completion dependencies exist shows limitations in case where task-completion dependencies interleave. 
The \emph{distributed objects} concept available to UPC++ programmers are equivalent to our \emph{local handles}.

Another approach close to ours is PCJ~\cite{pcj}.
This pure Java library brings a PGAS programming model to Java, relying on elegant annotations to mark the variables that belong to the global address space.
The library also provides collective communications operating on the variables of the global address space such as \verb|broadcast|, \verb|scatter|, \verb|reduce| and others~\cite{pcjcollective}. 
While close, the programming models employed by PCJ and APGAS for Java differ in that PCJ uses numbered ``threads'' as the main support for computation, with potentially multiple threads hosted within a single JVM.
In terms of program semantics, the PCJ threads would correspond to MPI ranks, but the collective communications between the threads are factorized by the supporting JVMs.
With the APGAS for Java library, the \verb|Place| abstraction strictly corresponds to a single JVM, with multiple asynchronous activities running in shared memory on the same process.
One advantage of PCJ over our approach lies in the fact that it supports these collective communications using a pure Java-based implementation while we rely on MPI. This means that PCJ is easily portable to non-traditional HPC infrastructures such as the cloud~\cite{pcjcloud}.

While the runtime we rely on combines APGAS for Java~\cite{apgasforjava} and MPI, we cannot consider it to be ``MPI+Apgas'' as we rely primarily on APGAS to manage code execution locality and termination. MPI is only used internally for specific communication patterns.
Our approach of a library to support parallel and distributed programs differs from approaches involving dedicated programming languages in that we make it possible for programmers to directly use any previous knowledge of a popular programming language Java.

\section{Conclusion and Future Work}
\label{sec:conclusion}

In this article, we presented our Relocatable Distributed Collections Library for the Java AGPAS programming model.
Our library allows users to write complex parallel and distributed programs by providing clear abstractions to handle both parallelism and distribution.

We established the programmability gains and the performance of our system using two well-known Java benchmarks.
Using the PlhamJ financial market simulator, we demonstrated the capability of programmers to balance the computational load between hosts using the integrated relocation mechanisms of our library.

The library we presented here will serve as the basis for several future works.
We are currently working on a load-balancer integrated with the library capable of relocating entries of a distributed collection as a distributed computation is taking place~\cite{integratedlb}.
Under this system, an action to perform on every element of the collection is given as a closure by the programmer and our library takes care of applying the given closure to every elements in the distributed collection, potentially relocating some entries along the way if load-unbalance occurs.

We did not cover topics related to resilience in this article.
Additions to the X10 implementation of the APGAS programming model have been made to this effect~\cite{resilientX10}, but they have yet to fully trickle down to their Java counterpart on which we rely on. 
We do plan to implement features that will allow programmers to easily backup the (distributed) state of their collections into checkpoints, making it possible to recover after a failure.

Finally, we are considering introducing support for elasticity to our library. 
Posner and Fohry recently demonstrated this possibility with the APGAS for Java runtime~\cite{apgasshrinkgrow}.
We believe our library would be a great help to programmers in such situations where the number of running processes increases and decreases over time thanks to the support for relocation features.
We already identified PlhamJ as an application that would benefit from such capabilities.

\begin{acks}
This work was supported by the JSPS KAKENHI Grants Number JP20K11841 and JP18H03232.
\end{acks}

\bibliography{bibli}

\appendix

\begin{figure*}
\begin{lstlisting}[caption={Main procedure of the PlhamJ distributed simulator},label={lst:plhamjwholecode},basicstyle=\small]
// Simulation-related collections
CachableArray<Market> markets;                                                      // market information
DistCol<Agent> agents;                                                                          // agents
DistBag<List<Order>> orderBag;                                              // orders submitted by agents
DistMultiMap<Long,AgentUpdate> contractedOrders;                           // trades contracted by agents
// Runtime variables
TeamedPlaceGroup world = TeamedPlaceGroup.getWorld();      // group of places involved in the computation
boolean isMaster = here() == place(0);                           // orders are handled by master=place(0)
long accumulatedOrderComputeTime = 0l;        // time spent on agent order-submission as part of (step 2)
int lbPeriod = 10;                                                  // load-balance period (configurable)
int iter;                                                                     // current iteration number

world.broadcastFlat(() -> {
  // (1) Broadcast the updated state of markets
  markets.broadcast(MarketUpdate::pack, MarketUpdate::unpack);
  // (2) Submit agent orders
  long startOrder = System.nanoTime();
  if (!isMaster) agents.parallelToBag((agent, orderCollector) -> {
      List<Order> orders = agent.submitOrders(markets);
      if (orders != null && !orders.isEmpty()) { orderCollector.accept(orders);}
    }, orderBag);
  accumulatedOrderComputeTime = System.nanoTime() - localSubmitTime;
  // (3) Collect all orders on the ''master'' place(0)
  orderBag.team().gather(place(0));
  // (4) Match buy and sell orders, populating `contractedOrders`
  finish(()->{
    // (4 - optional) balance the agents between places 1..n
    if (iter % lbPeriod == 0) {  async(()->{
        // Exchange time information between hosts
        long [] computationTime = world.allGather1(accumulatedOrderComputeTime);
        CollectiveMoveManager mm = new CollectiveMoveManager(world);               // prepare a relocator
        performLoadBalance(computationTimes, mm);               // various relocation strategies possible
        mm.sync();                                                              // perform the relocation
        accumulatedOrderComputeTime = 0l;                      // reset accumulated order-submission time
        agents.updateDist();                          // update the agents' distribution after relocation
      });
    }
    if (isMaster) handleOrders();                                    // details of this procedure omitted
  });
  // (5) Inform the agents of the trades they made
  // (5.1) Relocate contracted trade information to agents' location
  LongRangeDistribution agentDistribution = agents.getDistribution();
  contractedOrders.relocate(agentDistribution);
  // (5.2) Update the agents that contracted a trade
  if (!isMaster) contractedOrders.parallelForEach((idx, updates) -> {
      // Retrieve the agent targeted by the update
      Agent a = agents.get(idx);
      // Apply each update for this agent
      for (AgentUpdate u : updates) { a.executeUpdate(u);}
    });
});                                                                        // end of broadcast flat block
\end{lstlisting}
\Description{}
\end{figure*}

\begin{figure*}
\begin{lstlisting}[caption={Hybrid MolDyn implementation},label={lst:moldyn}]
TeamedPlaceGroup world = TeamedPlaceGroup.getWorld();
LongRange particleRange = new LongRange(0, nbParticles);
CachableChunkedList<Particle> particles; // Init omitted
int Ndivide = 5; // Number of columns/lines into which the product pairs are split
long seed = 0; // Seed used to assign the tiles to hosts

world.broadcastFlat(() -> {
	// Replicate the particles across process
	if (world.rank() == 0) {
		particles.share(particleRange);	
	} else {
		particles.share();
	}
	
	// Prepare the interaction pairs
	RangedList<Particle> prl = particles.getChunk(particleRange);
	RangedListProduct<Particle, Particle> product = RangedListProduct.newProductTriangle(prl, prl);
	// Split interactions into tiles and assign them to hosts
	product = product.teamedSplit(Ndivide, Ndivide, world, seed);
	
	// Prepare an accumulator for the force computation
	Accumulator<Sp> acc = new AccumulatorCompleteRange<>(particleRange, Sp::newSp);
	
	for (i = 0; i < iter; i++) {
		// Compute the force contribution of each pair
		product.parallelForEachRow(acc, (Particle p, RangedList<Particle> pairs, tla) -> {
			force(p, pairs, tla);
		});
		
		// Merge all the force contributions in the accumulators back into the designated particles
		particles.parallelAccept(acc, (Particle p, Sp a) -> p.addForce(a));
		
		// Sum the force contributions accross all hosts for each particle
		particles.allreduce((out, Particle p) -> {
			out.writeDouble(p.xforce);
			out.writeDouble(p.yforce);
			out.writeDouble(p.zforce);
		}, (in, Particle p) -> {
			p.xforce = in.readDouble();
			p.yforce = in.readDouble();
			p.zforce = in.readDouble();
		}, MPI.SUM);
		
		// Move the particles based on the computed force
		particles.parallelForEach(p -> move());
	}
});
\end{lstlisting}
\Description{}
\end{figure*}
\end{document}